# Image Features Influence Reaction Time:
# A Learned Probabilistic Perceptual Model for Saccade Latency

BUDMONDE DUINKHARJAV, New York University, USA
PRANEETH CHAKRAVARTHULA, Princeton University, USA
RACHEL BROWN, NVIDIA, USA
ANJUL PATNEY, NVIDIA, USA
QI SUN, New York University, USA

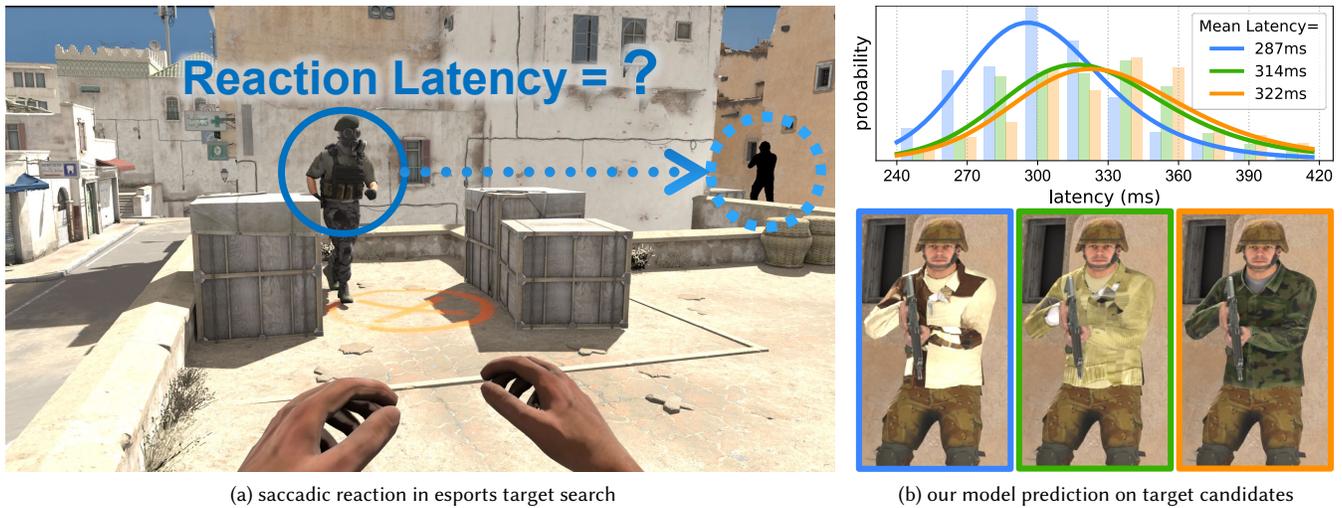

(a) saccadic reaction in esports target search

(b) our model prediction on target candidates

Fig. 1. *Illustration of our model and applications.* Our model predicts the reaction latency for users to identify and saccade to a peripheral target, as shown in (a). Based on our psychophysical data collected for stimuli with varying visual characteristics, we model the likelihood distribution of the time users take to process, react, and saccade to a target. If replacing the black placeholder in (a) with the three target candidates shown in (b), the resulting retinal images exhibit identical perceptual similarity in terms of visual acuity, with all FovVideoVDP scores >9.5 per [Mantiuk et al. 2021]. However, they may trigger significantly faster (leftmost of (b)) or slower (rightmost of (b)) reaction latencies with up to about 35ms difference, significantly affecting task performance. 3D asset credits to Slayver at Sketchfab Inc as well as Counter-Strike: Global Offensive © Valve Corporation.

We aim to ask and answer an essential question "*how quickly* do we react *after* observing a displayed visual target?" To this end, we present psychophysical studies that characterize the remarkable disconnect between human saccadic behaviors and spatial visual acuity. Building on the results of our studies, we develop a perceptual model to predict temporal gaze behavior, particularly saccadic latency, as a function of the statistics of a displayed image. Specifically, we implement a neurologically-inspired probabilistic model that mimics the accumulation of confidence that leads to a perceptual decision. We validate our model with a series of objective measurements and user studies using an eye-tracked VR display. The results demonstrate that our model prediction is in statistical alignment with real-world human behavior. Further, we establish that many sub-threshold image modifications commonly introduced in graphics pipelines may significantly alter human reaction timing, even if the differences are visually undetectable. Finally, we show that our model can serve as a metric to predict and alter reaction latency of users in interactive computer graphics applications, thus may improve gaze-contingent rendering, design of virtual experiences, and player performance in e-sports. We illustrate this with two examples: estimating competition fairness in a video game with two different team colors, and tuning display viewing distance to minimize player reaction time.

Authors' addresses: Budmonde Duinkharjav, budmonde@gmail.com, New York University, USA; Praneeth Chakravarthula, cpk@cs.unc.edu, Princeton University, USA; Rachel Brown, rachelbrown347@gmail.com, NVIDIA, USA; Anjul Patney, anjul.patney@gmail.com, NVIDIA, USA; Qi Sun, qisun@nyu.edu, New York University, USA.

# 1 INTRODUCTION

Measuring, modeling, and predicting how humans perceive and act on displayed visual content are important tasks in computer graphics, with applications in cinematic, real-time rendering, virtual/augmented reality (VR/AR), display optimization, esports, video compression/streaming, and visual design [Dunn et al. 2020; Mantiuk et al. 2004; Patney et al. 2016; Serrano et al. 2017; Sitzmann et al. 2018]. Perceptual image quality metrics predict the likelihood of visibility of image artifacts that result from creative and technical design, or are a side-effect of rendering, processing, or transmission. While many such metrics already exist, research is primarily focused on modeling the spatial/temporal *acuity* of the human visual system (HVS), not on how viewers "react" *after* seeing the stimuli. Although visibility may be closely related to behavior, there is evidence that perceptually identical stimuli frequently result in significantly different reactions for viewers [Mulckhuyse and Theeuwes 2010; Spering and Carrasco 2015]. Since responses are critical in many interactive applications such as esports and user interfaces, metrics that predict user reactive performance are arguably in emerging and crucial demand.

Researchers have so far exhaustively studied the acuity of the human visual system and established a significant body of perceptual image-quality metrics [Hore and Ziou 2010], as well as perceptually-optimized computer graphics techniques [Krajancich et al. 2021; Patney et al. 2016]. Such methods have unlocked significant performance and memory optimizations, as well as quality improvements. A recent example is gaze-contingent rendering, which perceptually optimizes rendering complexity for wide field-of-view (FoV) AR/VR displays. The rendering methods achieve great savings in computation [Patney et al. 2016], bandwidth [Chen et al. 2022; Kaplanyan et al. 2019; Krajancich et al. 2021], or enhancing visual cues [Krajancich et al. 2020]. Researchers have also proposed perceptually-based metrics that predict the visibility of artifacts in a user's peripheral vision [Mantiuk et al. 2021; Tursun et al. 2019]. However, to our best knowledge, there is still very limited characterization of changes to human reaction times when observing naturalistic visual stimuli that are otherwise perceptually identical.

This paper proposes an analytical model for a user's reaction time as evidenced by their eye movements. Human eyes change visual fixation three to four times every second [Fabius et al. 2019] via rapid exploratory movements called saccades. Saccades allow for frequent shifts of attention to better understand one's surroundings and to localize objects of interest, e.g., potential dangers [Purves et al. 2008]. Saccadic reaction latencies, after the eye observes a stimulus, are closely tied to performance in a broad range of real-time applications. For instance, subtly (as low as 4ms [Kim et al. 2019]) altered saccade latency can significantly determine performance in competitive esports [Koposov et al. 2020]. Each saccade involves perceiving a stimulus, identifying the target [Lisi et al. 2019], sending oculomotor neural signals, and controlling the extraocular muscles to reorient the eyeballs. Due to these complex mechanisms, fully characterizing changes in saccade/fixation as a function of changes in visual stimuli remains an open problem in vision science and computer graphics.

Note that, unlike with visual quality metrics, both high and low visibility of a target could hypothetically induce a longer processing time for fine details or blurred content. That may lead to potential non-correlation between acuity and saccadic latency [Kalesnykas and Hallett 1994]. Therefore, we present a visual-oculomotor and probabilistic model of the correlation between gaze-contingent visual stimuli and the timing of humans' cognitive decisions to perform saccades. Our model adapts to different viewing tasks that involve reacting to a stimulus in the visual field. Psychophysical experiments demonstrate our model's statistical effectiveness and real-world applications such as customizing players' action timing, measuring esports competition fairness, and predicting performance with various display environments.

We first conduct a preliminary psychophysical study. The results reveal that peripheral artifact visibility fails to fully explain changes in saccadic reaction latency. By leveraging the data, we develop a probabilistic and closed-form visual-oculomotor model predicting the likelihood of human saccadic latency after observing a given visual stimulus. The model is neurologically inspired to depict a decision-making process and is further established via machine learning with our studied data. We validate our model using various evaluation data partitioning and several realistic visual search tasks. Our results demonstrate that meaningful changes in behavior may occur even without visible artifacts, establishing the importance of predicting user gaze behaviors. Our main contributions are:

- a series of psychophysical experiments characterizing the disconnect between visual acuity and eye movement latencies toward a target;
- a neurologically-inspired and closed-form probabilistic model of human reaction time, built to effectively predict saccade onset latency by learning from real user data;
- studies with natural images and complex tasks showing how subtle target appearance changes may significantly alter reaction performance, following our model's predictions;
- demonstrations of the model's additional applications to measuring competitive game fairness and optimizing player performance in esports.

In addition, we provide the full source code and all raw user study data to the community at www.github.com/NYU-ICL/gaze-timing.

# 2 RELATED WORK

We propose a model for eye motion based on the characteristics of the human visual system, and our primary applications lie in the area of interactive computer graphics. Our model is an alternative to metrics for visibility of image artifacts, which serve for similar applications. Hence, we begin by discussing state-of-the-art in peripheral image perception and corresponding image quality metrics, as well as recent advances in the area of gaze-contingent rendering. We then discuss existing models for saccadic gaze motion and their applications in computer graphics and related areas.

## 2.1 Human Vision and Image Quality Metrics

The HVS is a complex bio-system that perceives the visual world using a series of optical components, retinal photoreceptors, and neural structures, including the optic nerve and visual cortex. Almost all of these are highly non-uniform across the visual field. As a consequence, our peripheral vision exhibits numerous peculiarities—in addition to having significantly lower acuity than foveal (central) vision, it is known to have nonuniform acuity for detection vs. resolution tasks [Thibos et al. 1987b,a], high sensitivity for moving stimuli [McKee and Nakayama 1984] including high critical flicker-fusion frequency [Hartmann et al. 1979], and reduced color perception [Cohen et al. 2020; Noorlander et al. 1983].

Several models describe aspects of peripheral visual perception and predict the visibility of stimuli at various visual eccentricities. Many build on contrast sensitivity measurements across the visual field [Barten 1999; Cajar et al. 2016; Daly 1992; Kelly 1979]. While most metrics only consider spatial image characteristics [Rimac-Drlje et al. 2010; Rimac-Drlje et al. 2011; Wang et al. 2001], some recent developments also account for spatio-temporal perception [Krajancich et al. 2021; Mantiuk et al. 2021]. However, all these metrics for peripheral image quality only account for visibility of artifacts and not for change of user behavior that might exist between perceptually identical stimuli.

## 2.2 Gaze-Contingent Computer Graphics

Gaze-contingent rendering utilizes high-speed eye-tracking to identify the user's gaze locations and to modify the rendered images or display accordingly. Such techniques can often provide additional perceptual cues to improve immersion of displayed content, e.g., perceived parallax [Konrad et al. 2020], stereo depth perception [Krajancich 2020], and depth-of-field [Duchowski et al. 2014; Hillaire et al. 2008; Mauderer et al. 2014]. Leveraging the perceptual differences between foveal and peripheral vision also improves interactive computer graphics methods, which forms the basis for gaze-contingent foveated rendering and displays. Foveated methods reduce the complexity of graphics or display in the peripheral visual field, which results in an overall improvement of performance or perceived quality [Franke et al. 2021; Guenter et al. 2012; Kaplanyan et al. 2019; Koskela et al. 2019, 2016; Meng et al. 2018; Patney et al. 2016; Polychronakis et al. 2021; Sun et al. 2017, 2020; Walton et al. 2021; Weier et al. 2016].

Existing gaze-contingent and foveated rendering algorithms rely on how gaze location causes differences in spatiotemporal visual perception. Typically, these methods are evaluated on the noticeability of any visual artifacts they may introduce. Instead, we attempt to model gaze *behavior* changes, e.g., saccadic latency, *after* observing a stimulus. The reactive changes occur due to both subthreshold and suprathreshold differences in visual stimuli, which can be leveraged to predict or improve user performance in immersive visual tasks.

## 2.3 Models for Saccadic Eye Motion

Saccadic eye motion exhibits several unique qualities. We find limited studies in the literature that model the latency before the onset of a saccade. Further, they do not characterize image features in a way that may build a general-purpose metric. We discuss research that explains the mechanism for latency that leads to a saccade and build on this mechanism for our model.

First, the large, rapid gaze changes that occur during a saccade are ballistic and exhibit a predictable trajectory [Bahill et al. 1975;

Kowler 2011]. Saccade amplitude, velocity, and duration are nonlinearly related, and velocities of short saccades tend to have an asymmetric, bell-shaped velocity profile [Bahill et al. 1975]. This behavior lends to models that can characterize a saccade profile by only partially observing it, and with a sufficiently fast eye-tracker, can predict *where* an ongoing saccade will land via regression [Arabadzhiyska et al. 2017] or neural network [Morales et al. 2021]. It can also help improve the perceived latency of gaze-contingent computer graphics, as studied by Albert et al. [2017].

Second, during a saccade as well as for a short period after, the HVS experiences temporary perceptual blindness, known as *saccadic suppression*, and has been well-studied in literature [Burr et al. 1994; Diamond et al. 2000; Ibbotson and Cloherty 2009; Matin 1975]. Saccadic suppression naturally helps gaze-contingent graphics be tolerant to higher eye-tracking latencies [Albert et al. 2017], and has also been leveraged in VR redirected walking [Sun et al. 2018].

Third, saccadic eye movements have been shown to be inaccurate and often undershoot their target [Becker and Fuchs 1969; Deubel et al. 1982]. The magnitude of error in the landing position depends on the degree of uncertainty in the target location as well as sensory noise [van Beers 2007] and adaptation [Cotti et al. 2009]. Researchers have proposed to model this uncertainty based on visual characteristics of the target [Carpenter 2004; Lisi et al. 2019]. Such models could help predict as well as correct for saccade errors by reducing the frequency and size of missed and catch-up saccades. However, we are not aware of such practical application of these models in prior work.

Finally, another area of saccade behavior models is subtle gaze direction [Bailey et al. 2009], which seeks to stimulate saccades toward specific targets by modeling how the spatio-temporal characteristics of a stimulus may attract a user's gaze.

While the above studies thoroughly explain and utilize various saccadic behaviors, we are not aware of computational models of pre-saccadic reaction time as a function of changes in naturalistic stimuli. Predicting this latency is potentially valuable for a variety of applications, such as esports [Koposov et al. 2020].

Various research endeavors to explain the cognitive decision that precedes and results in a saccadic eye movement. Several neuroeconomic approaches describe a generic decision-making process where the perceptual measure of *visual evidence* is accumulated stochastically. Once a *threshold* amount of evidence is collected, a decision is made [Mazurek et al. 2003]. The diffusion model [Palmer et al. 2005; Ratcliff 1978] suggests that the stochastic process may be modeled as a random walk akin to Brownian Motion. Carpenter [1995] assumes the *rate* of accumulation of *visual evidence* as a random variable which stays constant throughout individual saccades, as verified in [Reddi et al. 2003]. The above mechanisms suggest that sub-threshold changes in a stimulus could non-trivially influence the latency of saccade onset, as has been verified by [Mulckhuyse and Theeuwes 2010; Spering and Carrasco 2015]. Further, since stronger stimuli accumulate faster than weaker ones, the amount of time required must be correlated with various visual characteristics [Bell et al. 2006; Carpenter 2004; Mahadevan et al. 2018] and task modalities [Yamagishi and Furukawa 2020]. There is also evidence that the latency is a function of the target eccentricity and demographic [Mazumdar et al. 2019] but exhibits a distinct trend from that of visual acuity [Kalesnykas and Hallett 1994]. Inspired by these discoveries, we develop an analytical and data-driven model for saccadic reaction latency, dependent on visual contrast, frequency, and target locations. With competitive scenarios such as esports, we further demonstrate how our model can guide virtual asset design toward optimizing player performance and competition fairness.

Table 1. *Specifications of the HTC Vive Pro Eye display used in our studies.*

| Feature | Value |
|---|---|
| Display Resolution | $1440 \times 1600$ pixels per eye |
| Refresh Rate | 90 Hz |
| Peak Luminance | 143 cd/m$^2$ |
| Field of View | 110° diagonal |
| Eye Tracker Frequency | 120 Hz |

## 3 PILOT STUDY: MEASURING SACCADIC LATENCY

We first conduct a psychophysical experiment with parameterized stimuli to observe and measure the correlation between image characteristics and the time it takes to process them in order to trigger a saccade, and whether/how the correlation differs from that of visual acuity. We anticipate the data collected from the participants to serve as the inspirational and statistical basis toward a closed-form predictive model.

*Setup.* The study was performed with an eye-tracked HTC Vive Pro Eye head-mounted display as shown in Figure 2a. The hardware details are specified in Table 1. During the study, participants remained seated and perceived stimuli through the stereo display. Before each experiment, a five-point eye-tracking calibration was applied on each individual.

*Participants.* The psychophysical study was performed with $n = 5$ participants (ages $22 - 28$, 3 female) with normal or corrected-to-normal vision. The participants were instructed to perform a series of two-alternative forced choice (2AFC) tasks for each trial. The experiment was conducted during a single session split into 10 blocks, with each block containing 225 trials, i.e., 11250 trials in total with all the participants. The procedure took around 2.5 hours for each participant, including breaks between blocks, a short training session preceding the experiment, and a debrief afterwards.

*Stimuli and Tasks.* Figures 2a and 2b illustrate the experiment procedure and stimuli. The task is to:

(1) fixate at the center of the display,
(2) when visible, identify the orientation (i.e., symmetry axis) of the Gabor pattern presented at some eccentricity in the visual field, and
(3) make a saccade either to a left or a right target based on the orientation of the Gabor pattern.

We include Gabor patches for all combinations of contrasts ($c = \{.05, .22, .53, 1.0\}$), frequencies ($f = \{.5, 1.0, 2.0, 4.0\}$ pixels-per-degree), and eccentricities ($e = \{0°, 10°, 20°\}$). Three conditions (with ($c, f, e$) values of (.05, 4.0, 10°), (.05, 4.0, 20°), (.22, 4.0, 20°)) were excluded due to the patches not being detectable by all participants. The eccentricity range was chosen to cover common scenarios since the human gaze does not typically go outside 10° from the center [Hatada et al. 1980], and most natural saccade sizes are less than 15° [Bahill 1975]. Unless otherwise specified, we use Weber contrast in all our experiments and as input to our model.

At the beginning of each trial, the participants fixated at a cross shown in the center of the screen. Once they successfully fixated on the cross, it disappeared and a pair of Gaussian patches appeared at 10° eccentricities to the left and right of fixation. These patches served as the target locations to which the participants would saccade to indicate their decision about the stimulus. After a small delay—chosen randomly between 300 and 500 ms to avoid learning effects—the primary stimulus (Gabor patch) appeared either at the center of the screen (eccentricity=0°), or at some eccentricity in the inferior peripheral vision (eccentricity=10° or 20°). We instructed the participants to identify whether the Gabor stimulus was oriented at a rotation of 45° clockwise from vertical, as shown in Figure 2a, or 45° counter-clockwise from vertical. We further instructed them to saccade to the target patch corresponding to their determination, right for clockwise and left for counter-clockwise. During each trial we recorded the subjects gaze at a rate of 120 FPS using the display's built-in eye tracker.

We varied the eccentricity, contrast, and frequency of Gabor patterns across trials such that each combination of variables was shown 5 times in each block for 10 blocks, yielding a total of 50 trials per condition. To ensure the participants were completing the task correctly, we discarded all trials where they do not complete the task correctly, and repeat all mistaken trials at the end of the block until all trials are completed. The order of these conditions was randomly shuffled within each block to eliminate any bias. Meanwhile, all features of the Gaussian target patches (only being used to cue the saccade direction) remained unchanged throughout the trials. For the practice session at the beginning of the experiment, each participant performed one block of the study with identical settings as in the actual study. Please refer to our supplementary video for an animated illustration.

### 3.1 Results

Using a statistical saccade detection method in Section 5.3, we identify the saccadic latency as the duration between appearance of the primary stimulus (Gabor patch) and the first frame of a participant's saccade. We notice that the saccade latencies exhibit an asymmetrical distribution as shown in Figure 2c. As the various features of the stimulus are modulated, the overall shape of the distribution remained consistent while the mean saccade latency varied by as much as 25% or 100 ms as shown in Figure 3. Increasing the contrast of the stimuli decreases reaction latency, while increasing the frequency increases the latency. Further, increasing the eccentricity does not always reveal a monotonic effect, but instead a U-shaped effect with the lowest mean latency values (265 ms) plateauing at 10°. For breakdown visualizations, please refer to Figure 3/Appendix E for the effects of individual characteristics and participants.

We also analyzed FovVideoVDP [Mantiuk et al. 2021] scores for all stimuli used in the experiment. Using the median condition's image ($c = 0.53, f = 2$ cpd, $e = 10°$) as reference, we observed $9.52 \pm .03$ (out of 10) value across all the stimuli being studied.

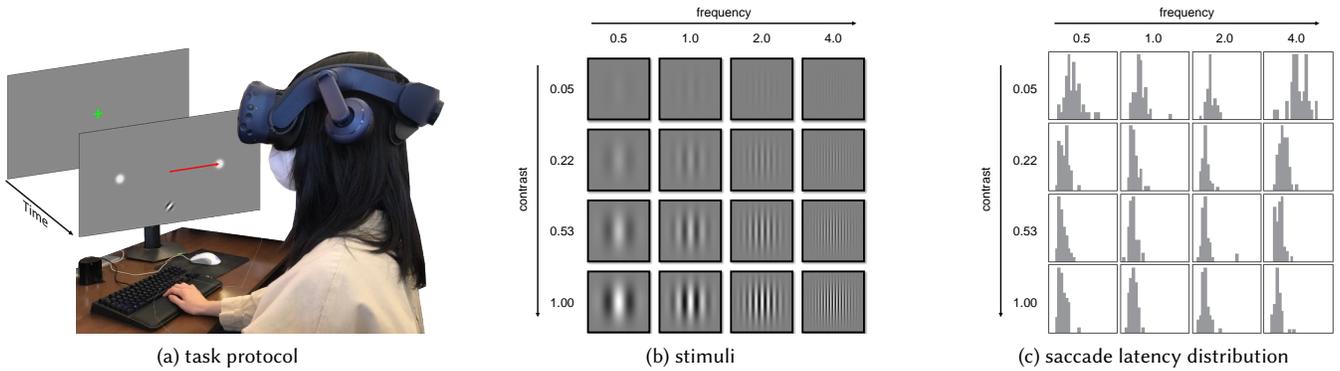

(a) task protocol    (b) stimuli    (c) saccade latency distribution

Fig. 2. *Preliminary user study procedures and results.* (a) shows our setup and the study procedure: two *target* Gaussian patches are shown left and right from the initial fixation. After a brief delay of 300 to 500 ms, a *reference* Gabor stimulus appears in the inferior periphery. If the *reference* stimulus is oriented at 45° clockwise from the vertical axis, the correct *target* saccade location is on the right side, and vice versa for a *reference* stimulus with the opposite orientation (i.e. counter-clockwise orientation). The *latency* of the saccade response indicating the decision is recorded. Across trials, the contrast, frequency, as well as vertical eccentricity of the *reference* Gabor stimulus are varied as experimental parameters. *Target* Gaussian patches are unchanged across all trials. (b) visualizes all the stimuli used for this study. Chosen contrast values are $c = \{0.05, 0.22, 0.53, 1.0\}$ as measured by Weber contrast; frequency values are $f = \{0.5, 1.0, 2.0, 4.0\}$ cpd. All stimuli were shown at eccentricity values of $e = \{0°, 10°, 20°\}$. (c) histograms of saccade latencies for one sample subject when the reference stimulus was located at 0° eccentricity. The distributions exhibit a skewed asymmetrical shape, similar to other distributions of reaction time in related work (see Section 2.3). With $\{c = 0.53, f = 2 \text{cpd}, e = 10°\}$ as the reference stimulus, all stimuli images (from (b)) show high and similar FoVVDP scores ($9.52 \pm 0.03$), despite significant variances in their resulting saccade latencies.

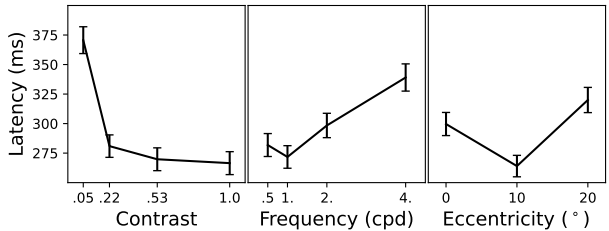

Fig. 3. *Aggregate trends of our preliminary study dataset.* The pilot study raw data is aggregated using either contrast, frequency or eccentricity of the reference Gabor patch, and averaged across the other two variables. Error bars represent standard error of measurement. Reaction times decrease as visibility of the stimuli is improved, and vice versa. Surprisingly, the reaction latency when the stimulus is at the fovea is higher as compared to when it is in mid-periphery.

### 3.2 Discussion

The above results and analysis reveal several remarkable discoveries on the relationships between visual characteristics and saccadic latency. First, using a state-of-the-art peripheral image similarity metric [Mantiuk et al. 2021], we conclude that visual differences between our stimuli are all well below the perceivable thresholds. Yet, they result in a significant difference in saccadic latency. This evidence confirms that perception of visual differences cannot alone explain the changes in saccadic latencies. Second, the asymmetrical probability distribution agrees with the discoveries of prior work in measuring similar visual-oculomotor reactive latencies [Carpenter and Williams 1995; Lisi et al. 2019; Palmer et al. 2011]. Third, that at a given eccentricity, as the visibility of the stimuli improves (either by increasing contrast or by modulating the frequency), the latency decreases. Meanwhile, the latency rises toward infinity whenever visibility reduces and approaches the Contrast Sensitivity Function (CSF) threshold. Lastly, we observe a surprising effect that the saccade latencies for a stimulus at the fovea are longer than in mid-periphery. We hypothesize that the more analytic purpose of the fovea causes feature extraction to take longer, similar to the results reported by Kalensnykas et al. [1994].

The collected data and the observations drive our development of a closed-form probabilistic model inspired by the computational process of decision making, as detailed in the next section.

## 4 PROBABILISTIC MODEL OF SACCADIC REACTION

Driven by our preliminary study discoveries, we aim to establish the computational relationship between saccadic reaction latencies and visual characteristics of stimuli. In Section 4.1, we model the saccadic behaviors as a random process of decision-making and incorporate perceptual, cognitive, and individual uncertainties. Section 4.2 shows how image characteristics such as contrast, frequency and eccentricity relate to our model of saccadic onset latency for a task. Sections 4.3 and 4.4 detail how we leverage our preliminary study data to learn the resulting parameters for our model.

### 4.1 Random Process Model for General Decision-Making

The Drift Diffusion Model (DDM) has been leveraged in neuro-economics and psychology to model decision-making behaviors in perceptual tasks [Fudenberg et al. 2020]. Using this model, we may quantify the reaction latencies for performing various tasks such as to decide whether to stop a car upon seeing an approaching object, or to correctly identify a friend for in a video game. Throughout a decision process, DDM presents a measure of "evidence" which is used to quantify how much confidence an individual needs to reach a decision. "Evidence", in this context, is accumulated over time, and once the amount of evidence reaches a minimum threshold, a decision/action is triggered, as visualized in Figure 4. As soon as a decision is made, the evidence is reset, and the process restarts, preparing for the next decision.

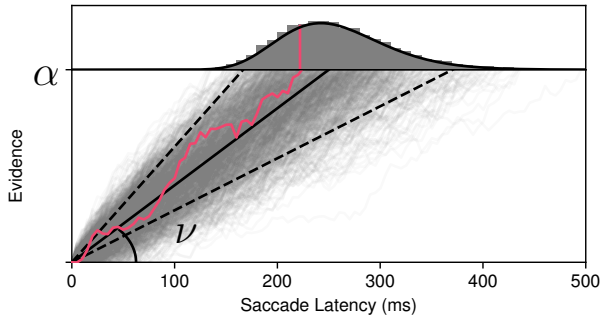

Fig. 4. *Visualization of our stochastical model.* The x-/y-axis indicates time/perceptual evidence levels. With a task-determined evidence threshold $\alpha$ and the image-determined accumulation speed $\nu$, each saccade decision is modeled as a random walk process (for instance, the pink curve). When the evidence accumulation reaches $\alpha$, a saccade is triggered. Due to cognitive noise and individual differences, each saccadic decision may vary (as in the light gray curves), thus forming the action timing as a probabilistic event. Please refer to our complimentary video for a dynamic visualization.

The process of evidence accumulation is modelled as a random stochastic process to reflect human reaction uncertainties and inconsistencies. As the name of the DDM suggests, the observed evidence is modelled as a diffusion process with non-zero drift, also known as Brownian motion.

*Integration-and-action.* We represent the measure of evidence accumulation after time $t$ via a stochastic process, $\{A(t; \nu)\}_{t \geq 0}$. The process is modelled as the trajectory following Brownian motion with mean drift rate $\nu$. Formally, the process is expressed as

$$A(0; \nu) = 0,$$
$$A(t; \nu) = \nu t + W(t) \qquad (1)$$
$$W(t) \sim \mathcal{G}(0, t),$$

where $\mathcal{G}$ is the Gaussian distribution.

As illustrated in Figure 4, at the beginning of a decision-making process, no evidence has been accumulated to affect the confidence of reaching a decision. Therefore, the initial evidence amount, $A(0; \nu)$, equals to zero. As time, $t$, progresses, confidence builds up with a mean rate of $\nu$, while also accumulating noise proportional to $t$, due to various uncertainties in the process of evidence accumulation.

The distribution of *evidence*, $A(t; \nu)$, at a given point in *time*, $t$, can be simplified and expressed as a *Gaussian* distribution,

$$A(t; \nu) \sim \mathcal{G}(\nu t, t). \qquad (2)$$

However, we aim to characterize the distribution of the *latency* when enough evidence is collected to trigger a decision. More concretely, we seek to find the distribution of the earliest *time*, $T(\alpha; \nu)$, when the integrated evidence reaches a given threshold, $\alpha$:

$$T(\alpha; \nu) := \inf_t \{A(t; \nu) = \alpha\}. \qquad (3)$$

This measure of earliest reach to the threshold, $T(\alpha; \nu)$, corresponds to the reaction latency of the decision being made.

Solving for $T(\alpha; \nu)$ using Equations (1) and (3) (see Appendix A for derivation), we find that $T(\alpha; \nu)$ follows the *Inverse Gaussian* (IG), or Wald, distribution [Folks and Chhikara 1978]:

$$T(\alpha; \nu) \sim \mathcal{IG}(\alpha, \nu), \qquad (4)$$

which has a probability density function of

$$h(t; \alpha, \nu) = \frac{\alpha}{\sqrt{2\pi t^3}} \exp \frac{-(\alpha - \nu t)^2}{2t}. \qquad (5)$$

Intuitively, random variables for evidence, $A(t; \nu)$ and latency, $T(\alpha; \nu)$, roughly describe inverse functions of the same stochastic process, and fittingly follow Gaussian, and Inverse Gaussian distributions respectively.

### 4.2 Saccade as a Visual-Oculomotor Decision-Making

Equation (4) depicts generic decision-making latencies, parameterized with the evidence threshold, $\alpha$, and mean evidence integration rate, $\nu$. A saccade is a representative example of visual-oculomotor decisions. A saccade latency results from processing visual stimuli to gain evidence of the target, and then deciding to make an eye movement. In this section, we determine how these two parameters in Equation (4) relate to the characteristics of visual content, such as contrast, frequency, and eccentricity, as well as the nature of the visual task of interest.

*The evidence threshold*, $\alpha$, is primarily dependent only on the nature of the visual task. That is, what a visual task is asking an individual to execute determines how confident the individual needs to be in their belief to successfully make a saccadic decision. For instance, Palmer et al. [2011] show that $\alpha$ is invariant to modulations in signal strength of the visual stimuli but varies *across* different tasks such as feature search, conjunction search, and spatial configuration search. In fact, Reddi et al. [2003] demonstrate that if an individual is instructed to observe the same visual stimulus while executing different tasks, the evidence integration rate, $\nu$, stays constant while the evidence threshold, $\alpha$, varies between the tasks.

We thus correlate $\alpha$ with a visual task via a *task description*, $D \in \mathcal{T}$, where $\mathcal{T}$ is the set of all visual task descriptions which trigger viewers' saccades toward a target. Examples include searching for characters of the opposing team in a given esports game, comparing and choosing a preferred target, or the task described in Section 3.

*The evidence integration rate*, $\nu$, is shown to change depending on the difficulty to process the visual content [Palmer et al. 2011]. Results from our preliminary study (Section 3.2) reveal that visual characteristics have a complex, and non-monotonic relationship with processing difficulty of the stimulus. The effects are also, to some extent, naturally uncertain due to motor/neural noise and individual variances.

These results motivate us to model $\nu$ as a function of contrast ($c$), frequency ($f$), and eccentricity ($e$) of the target stimulus—some of the most important features which affect our perception of visual stimuli. The decoupled nature of the task and visual parameters, therefore, allow for modular computation of each parameter independently to derive a saccade-tailored decision latency expression $T_{sac}$ from

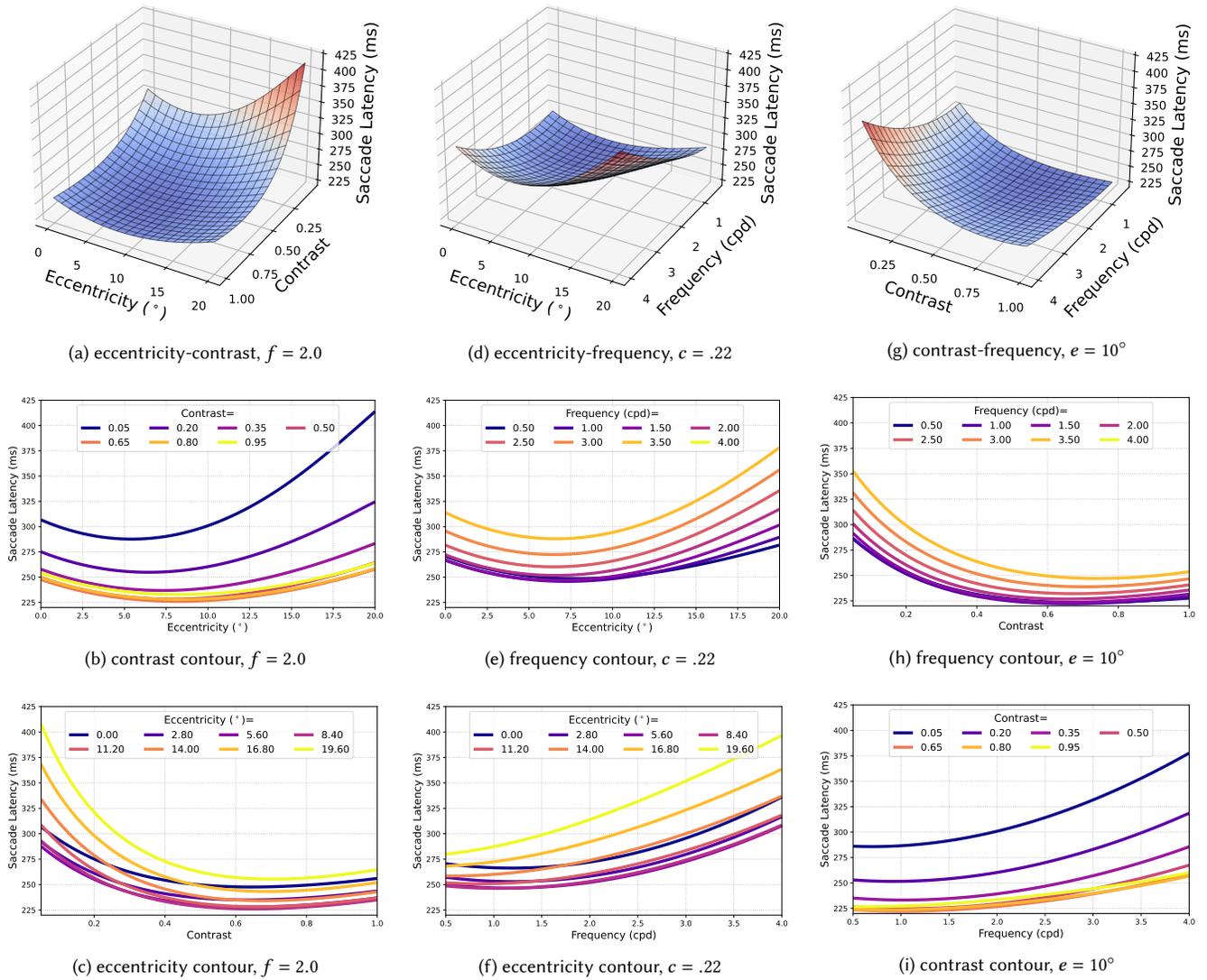

(a) eccentricity-contrast, $f = 2.0$

(d) eccentricity-frequency, $c = .22$

(g) contrast-frequency, $e = 10°$

(b) contrast contour, $f = 2.0$

(e) frequency contour, $c = .22$

(h) frequency contour, $e = 10°$

(c) eccentricity contour, $f = 2.0$

(f) eccentricity contour, $c = .22$

(i) contrast contour, $e = 10°$

Fig. 5. *Visualization of our model.* With a given task $D$, our model, defined in Equation (6), is $\mathbb{R}^3 \rightarrow \mathbb{R}$. The first row visualizes each two of the three dimensions $(c, f, e)$ as the variable to the saccade latency (z-axis). The second/third rows are the corresponding contours created by projecting the model to x-z/y-z axes. Note the U-shaped effects of $e$, and the inverse effects between $f$ and $c$.

Equation (4):

$$T_{sac}(D, c, f, e) \sim \mathcal{IG}(\alpha(D), \nu(c, f, e)). \quad (6)$$

In the following sections, we determine how to obtain the analytical forms of $\alpha(D)$ and $\nu(c, f, e)$ respectively by leveraging our collected data. Further, note that, as intuitively visualized in Figure 4, $\nu$ already universally determines the *relative* proportion between the mean and variance of latencies. It depends on the stimuli characteristics, and together with task-dependent $\alpha$ that determines the absolute time.

### 4.3 Measuring Evidence Threshold $\alpha$ in eq. (6)

To fit our model to the pilot user study data from Section 3, we need to compute $\alpha$ for the task $D_{pilot}$, and also fit a function which maps $(c, f, e)$ values to corresponding $\nu$ values.

Since our model is built with a single task description, denoted as $D_{pilot}$, and is kept invariant throughout the entire study, we only evaluate a single value of $\alpha$. We leverage how the mean, $\mathbb{E}$, and variance, $\mathbb{V}$ of the IG distribution relate to its $\alpha$ parameter to derive an estimation given a sample of saccade latency $T_{sac}(D_{pilot}, \cdot)$,

$$\hat{\alpha}(D_{pilot}) = \sqrt{\frac{\mathbb{E}[T_{sac}(D_{pilot}, \cdot)]^3}{\mathbb{V}[T_{sac}(D_{pilot}, \cdot)]}}. \quad (7)$$

Note that $\alpha$ only depends on the task but not the stimuli. Therefore, we can choose any $(c, f, e)$ in order to compute a $\alpha$. Extending the model from the pilot study task to novel tasks (e.g., stimulus comparison) only requires a calibration via Equation (7) using a sample drawn from the novel task.

### 4.4 Learning Evidence Integration Rate $v$ in eq. (6)

With the task-calibrated $\alpha$, we optimize the evidence integration rate, $v$ via maximum likelihood estimation (MLE) given a combination of stimulus' contrast, frequency, and eccentricity, $\{c, f, e\}$. Note that since the task description, $D$, does not correlate with $v$, the learned result is applicable to any arbitrary scenario and task. Mathematically, we formulate $v$ as a function of $\{c, f, e\}$. To ensure local smoothness of the desired function within this input domain, we model the function as a Radial Basis Function Neural Network (RBFNN),

$$v(c, f, e) = \sum_{i=0}^{N} \lambda_i \rho \left( \left\| \begin{bmatrix} c \\ f \\ e \end{bmatrix} - \mathbf{b}_i \right\|, \sigma_i \right), \tag{8}$$

where $\mathbf{b}_i$ indicates the individual radial basis centers, and $\rho$ is a Gaussian Basis function. In our experiments, we choose $N = 20$. Using our collected data from Section 3, we jointly train the RBFNN's weights $\lambda$, centers $\mathbf{b}_i$, and Gaussian deviations $\sigma$.

We need $v$ label values from our user study to obtain the RBFNN parameters. Similar to Section 4.3, we utilize the relationship between the evidence integration rate, $v$, and the mean, $\mathbb{E}$, of the IG distribution:

$$\hat{v}(c, f, e) \sim \mathbb{E}[T(\cdot, c, f, e)]^{-1}. \tag{9}$$

The proportionality constant of this relationship depends on the unit of time measurement and is set to 1 for simplicity, and scaled to appropriate units as necessary.

Detailed learning implementation can be referred to in Section 5.4. The final fitted model is visualized in Figure 5. Eccentricity's U-shaped effects on saccadic latency can be observed.

## 5 IMPLEMENTATION

### 5.1 Data Normalization

While saccade latencies vary according to the trends observed in Section 3, the absolute values measured across different individuals can vary significantly. In order to aggregate the data efficiently without introducing large amounts of noise caused by individual variances, we normalize the data as a first step during analysis. Specifically, we pick a calibration condition from all the data collected from a single experiment block and set the mean normalized duration of this condition equal to 1. For the pilot experiment from Section 3, the calibration condition was $c, f, e = \{1.0, 1.0, 0.0\}$. For the dual task experiment from Section 6.3, the calibration condition was $c_f, c_p = \{1.0, 1.0\}$. For the natural task experiment from Section 6.2, the calibration condition was the control group. The choice of all calibration conditions are arbitrary and we validated that it does not affect the model's predictive ability. See Appendix B for detailed description of normalization.

### 5.2 Hardware and Stimuli Generation

All user study systems were implemented in the Unity Game Engine and run on the HTC Vive Pro Eye with specifications in Table 1. The synthetic image datasets used in Section 6.2 are generated using 3D-assets purchased from the Unity Asset Store and rendered using the Cycles Rendering Engine included in Blender. When evaluating the FoVVideoVDP scores of the stimuli used in our work, we used the authors' open-source implementations which provided configurations for the HTC Vive Pro Eye. Some visualizations used in the demo video were created using the Manim Mathematical Animation Framework [2022].

### 5.3 Saccadic Latency Detection

Our method of detecting reaction times for saccadic events is measured by the time of onset of the "primary" saccade that is used to move the gaze to the target location. We define the "primary" saccade as the saccade that is onset and offset within $3°$ of the intended gaze origin and target locations respectively. For saccade detection we use the method presented by Engbert and Mergenthaler [2006]. Note that this saccade detection algorithm is not just limited to microsaccade detection as the title might suggest, and has been used by Lisi et al. [2019] for normal saccade detection.

### 5.4 Integration Rate Learning

To train the rate parameter, $v$ in Section 4.4, we train a Radial Basis Function Neural Network (as formulated in Equation (8)) implemented using the auto-differentiation library, Pytorch. We use the Adam optimizer with a learning rate of 0.1. The training used 2000 epochs and took $\sim 180$ seconds on a single NVIDIA RTX 3090 GPU.

## 6 EVALUATION

We gauge the ability of our model to generalize and consider possible applications via a series of psychophysical and simulated experiments. In Section 6.1, we first measure robustness and generalizability using data collected in Section 3. Next, in Section 6.2 we provide three example tasks featuring complex stimuli (athletics, esports, and photographic scenes), and compare the model's predictions with our user study data. In Section 6.3, we evaluate how our model extends to more sophisticated tasks containing multiple stimuli of interest. Finally, we present two possible applications of our model: evaluating esports competition fairness, and estimating how human-display configuration can alter in-game target searching performance in Section 7.

### 6.1 Model Performance and Generalizability

We present an analytical model whose parameters are learned based on user-collected data. Thus, we need to evaluate our model's performance regarding both prediction accuracy and generalizability beyond the specific trials and subjects included in the training data. To this end, we perform a comparative analysis of alternate training and testing partitions of the dataset.

*Protocol.* For each analysis, we reserve a different partition of the dataset from Section 3 for testing, re-train the model in the same way as in Sections 4.3 and 4.4, and then measure the accuracy of the re-trained model prediction on the reserved test data. We perform two types of partitioning protocols for reserving the test set:

(1) Random: a random selection drawn from all data points (20%),
(2) Subject: all data from each individual subject (20%).

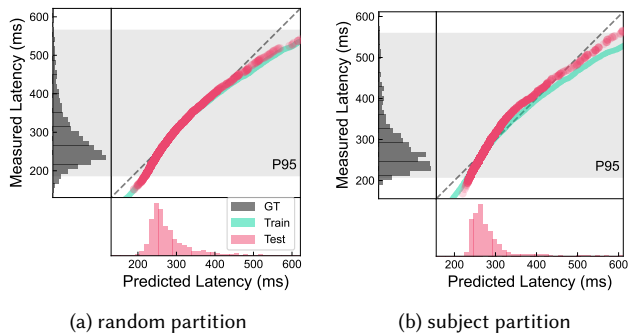

(a) random partition          (b) subject partition

Fig. 6. *Model performance and generalization validation using preliminary user study dataset.* Q-Q plots for various train/test splits demonstrate our model's accuracy and ability to generalize when applied to new data. In each plot, we visualize a comparison of histograms between the ground truth data (gray), and the model predictions on the test dataset (magenta). We also visualize the baseline Q-Q plot for the model's performance on the training dataset (in teal). The P95 confidence interval is highlighted in each figure to contextualize the volume of data being shown. As defined in Section 6.1, (a) shows the results with random partition (1). (b) shows the results with subject_01's data partition (2). We observe a distribution agreement between our model prediction and the unseen testing dataset; the closer the Q-Q curves are to the diagonal line, the more accurate the predictions are. The corresponding K.S. tests evidence the observation.

*Metrics and results.* We perform the Kolmogorov–Smirnov (K.S.) goodness-of-fit test between the reserved test data and our prediction [Massey Jr 1951], and show the Quantile-Quantile (Q-Q) plot [Gnanadesikan and Wilk 1968] in Figure 6. A significant difference on the K.S. test indicates a *rejection* of the null hypothesis that the sampled data is drawn from the same distribution; failing to reject the null hypothesis supports the accuracy and generalizability of our model. The Q-Q plot visualizes the correspondence of two probability distributions at each quantile. . Data below the $y = x$ line in Figure 6 indicate an overestimation of saccade latencies and vice versa for data above the line.

Figure 6 shows the Q-Q plot for the training and testing sets across both partition protocols. The K.S. test fails to reject the null hypothesis that the observed user saccade latency distribution is drawn from our model-predicted distribution for (1) the random partition, $D = .2$, $p = .99$, and (2) the individual subject partitions:

| Subject ID | S1 | S2 | S3 | S4 | S5 |
|---|---|---|---|---|---|
| K.S. analysis | $D = .3$ $p = .79$ | $D = .2$ $p = .99$ | $D = .2$ $p = .99$ | $D = .2$ $p = .99$ | $D = .1$ $p = 1.0$ |

where $D$ is the distance metric between two CDFs according to the K.S. Test, and $p$ is the $p$-value corresponding to the distance metric.

*Discussion.* The above analysis demonstrates that our model does not predict statistically different distributions compared to unseen observations across various partitioning protocols. The results of the randomly partitioned study (1) demonstrate the generalizability of our model across trials without observed overfitting. Analysis of the subject-partitioned study (2) verifies our model's applicability to unseen users, and thus general human saccadic behaviors.

We further performed an ablation study with individual visual characteristic conditions ($\{c, f, e\}$) provided in Appendix F. It compares each condition's contribution to our model by training an ablated version of the model with various conditions of the dataset missing, and quantifying the size of the regression in the model's predictive ability. The results show that across ablation conditions, the Mean Squared Error (MSE) between ground truth and our model's prediction regresses by 50% on average.

### 6.2 Predicting Saccadic Behaviors with Altered Target Appearances in Natural Tasks

In Section 3, we observed that unnoticeably subtle visual changes may induce significantly varied reactive latencies, as was formulated and predicted by our model in Section 4. In this experiment, we evaluate our model's application in several realistic target search scenarios such as esports, and real-world photographs.

Via a series of psychophysical studies, we seek to determine: (1) whether our model can extend to predicting saccadic reaction latencies with natural task/stimuli; (2) whether we can imperceptibly alter the appearance of objects while still introducing enough reactive latency to materially influence real-world task performance. We answer these queries in our experiment and compare our findings to the model predictions.

*Participants and setup.* We recruited 14 participants (ages $22 - 33$, 3 female) for this series of 2AFC experiments. Two participants were excluded for inability to perform the tasks (self-reported difficulty perceiving peripheral stimuli and target identification accuracy greater than one standard deviation below the mean), resulting in 12 final participants. Two of the 12 participated in the preliminary study in Section 3. The study was conducted during a 10-minute sessions consisting of 153 trials per scene for each participant. The hardware and setup remain the same as in Section 3.

*Scenes and stimuli.* To simulate a broad range of applications, our user study stimuli consisted of three groups of images: (1) a synthetic soccer scene, (2) a synthetic first-person view as an analog for esports, and (3) digital photographs of an indoor shelf. Each group contained 51 different images; each has the target stimuli appearing at different locations (to avoid learning effects) on the visual field, and serve as a trial. The background and targets from each evaluation group are shown in Figure 7. Although shown in color in the paper for visual clarity, all images were rendered with grayscale on display to avoid bias from color cues.

*Tasks.* Participants were instructed to complete a similar 2AFC decision task across all trial. At the beginning of each trial, they were shown a background image containing several task-irrelevant objects. After a randomized $1 - 1.5$ second delay, an additional task-relevant stimulus, either a *target* or *non-target*, appeared on the scene as in Figure 7. Participants were shown both types of stimuli ahead of the experiment. The task was to saccade to targets, or remain fixated if the stimulus was identified as a non-target. This procedure allows us to measure the visual-oculomotor latency after which a subject identifies the discernible feature of interest from the stimulus. This emulates the common real-world scenarios where a new "intruder" of potential interest enters the subjects' visual fields. Please refer to our video for dynamic illustrations of the task.

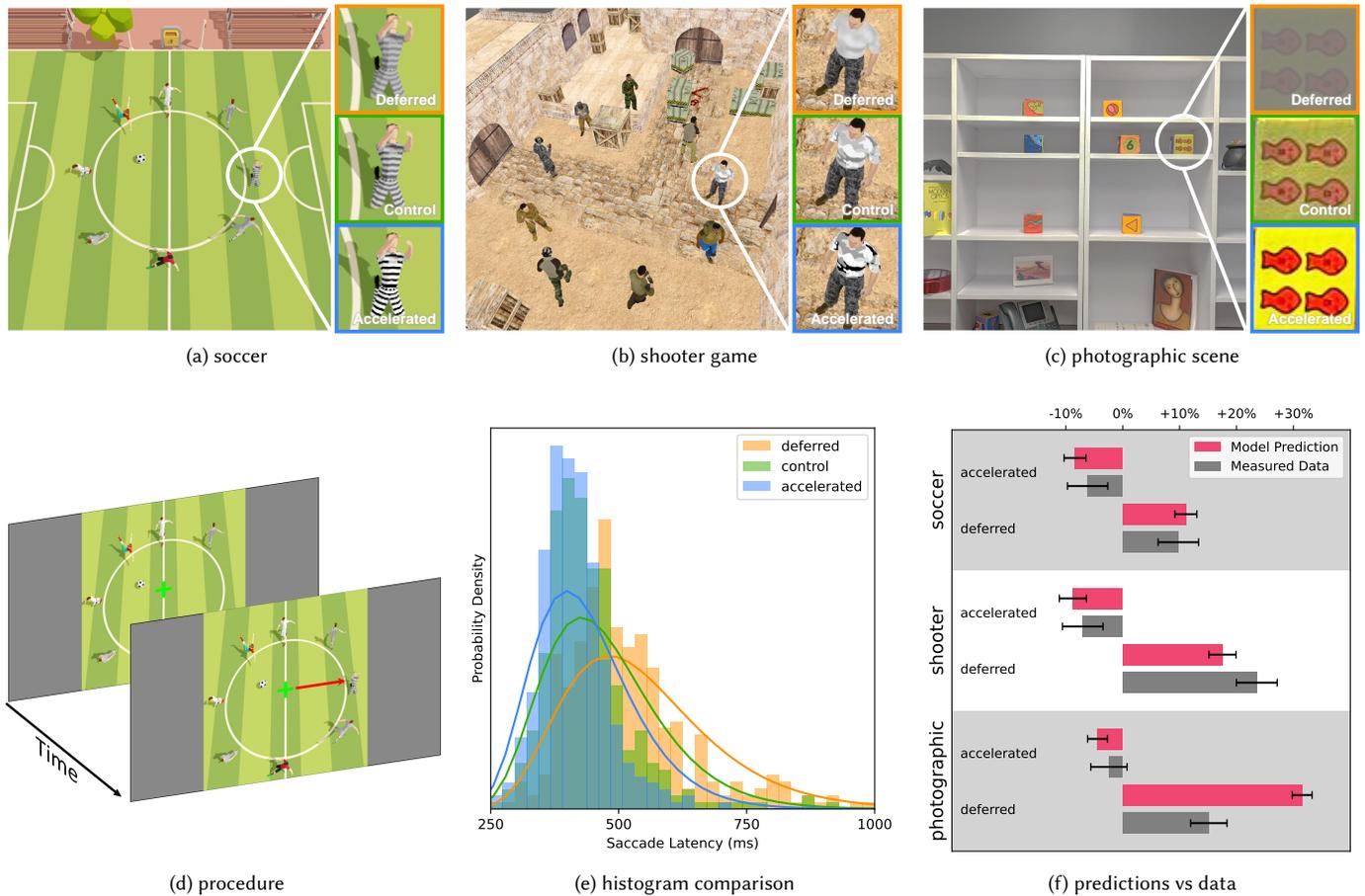

(a) soccer      (b) shooter game      (c) photographic scene

(d) procedure      (e) histogram comparison      (f) predictions vs data

Fig. 7. *User data and our model prediction of saccade latencies among varied target appearances.* Saccade latency modulation correlates with the contrast of stimuli as shown in the three distinct scenes (and target candidates) shown in (a)/(b)/(c). Each scene presents distinct visual characteristics including low polygon 3D scenes, dense geometries, or natural scenes. (d) illustrates the study procedure over time. With the **Control** condition as reference, all others show FovVideoVDP scores above 9.5, indicating identical perceptual appearance per [Mantiuk et al. 2021]. Using the shooter scene as example, (e) shows the user latency data in histograms, and our model predicted latency in curves. A significant agreement can be observed. Please refer to our supplementary videos for an animated visualization. (f) shows the mean relative durations (with **Control** as "0%" pedestal) of **Deferred**/**Accelerated**. The error bars indicate standard error. Full statistical analysis on all scenes can be seen in Section 6.2. Each individual's raw probabilistic distributions are provided in Appendix G. 3D asset credits to haykel-shaba (a), and Slayver (b) at Sketchfab Inc.

*Conditions.* Across each image set, we tested three variations of the target stimulus in order to measure how changes in image features affect saccade latencies. In one variation the target stimulus had increased contrast and/or decreased frequency (**Accelerated**), in another variation the target had decreased contrast and/or increased frequency (**Deferred**), and a third unfiltered variation was used as a control group (**Control**). Each participant performed 51 images × 3 conditions × 3 scenes, resulting in 459 trials total, i.e.,

5508 trials across the experiment. Measuring the precise frequencies affecting saccade latency is a complex task requiring pooling from multiband. Investigating a comprehensive pooling strategy is beyond the scope of this work. Therefore, we approximate the representative frequency as the Laplacian pyramid layer with the highest corresponding contrast, for those images without a uniform frequency pattern. Contrast and eccentricity computations were trivial to compute without requiring pooling operations.

*Results.* We present the results of our experiments in Figure 7. We again use the K.S. statistical test to evaluate alignment between predicted and measured histograms across the different scenes for each condition. We report the results of these tests below:

| | Deferred | Control | Accelerated |
|---|---|---|---|
| Soccer | $D = .2, p = .99$ | $D = .2, p = .99$ | $D = .2, p = .99$ |
| Shooter | $D = .2, p = .99$ | $D = .3, p = .79$ | $D = .1, p = 1.0$ |
| Photographic | $D = .3, p = .79$ | $D = .1, p = 1.0$ | $D = .2, p = 1.0.$ |

Please refer to Appendix G for the collected saccadic latency distributions of individual participants and scenes.

Using the **Control** images as reference, we additionally calculate the FovVideoVDP values for all images in our dataset. We find the mean values to be above 9.5 for all **Accelerated/Deferred** images, which indicates observers would be approximately at chance for detecting differences between them.

We also debriefed each participant after the experiment on their thoughts regarding the tasks, and most participants reported no self-awareness of reaction time difference.

*Discussion.* Our results demonstrate agreement between the predictions made by our model and the observed saccadic latency distributions across 12 participants. We find significant differences in saccadic latency across conditions, despite identical perceptual appearance evidenced by the FovVideoVDP metrics.

Our prediction of the photographic scene results show correct trends and distribution ratios, albeit for a scaled absolute time (in ms) relative to the measured data. We attribute this scale variance to the fact that natural images contain wide frequency bands and our single-frequency pooling in the Laplacian Pyramid may discard significant frequency information. This motivates interesting future work on multi-frequency pooling models tailored for reaction time, see Section 8.

### 6.3 Extending to Foveal-Peripheral Dual Tasks

In various real-world scenarios, humans perform tasks by jointly analyzing both foveal and peripheral content, such as with reading, film watching, and architectural design. In this experiment, we extend and evaluate our model to such applications considering *dual* tasks.

*Modeling.* Our visual system processes foveal and peripheral stimuli independently and in parallel for a variety of tasks [Ludwig et al. 2014]. That is, the foveal and peripheral pathways gather information concurrently, and the decision to trigger a saccade waits until both processes have finished. We hypothesize that these independent foveal and peripheral stimulus processing units operate using the *integration-and-action* process as described in Section 4.1.

In this model, processing times for both the fovea, $T_f$, and periphery, $T_p$, follow Equation (4), and can be adapted to specific task descriptions and stimulus characteristics as shown in Equation (6):

$$T_f \sim \mathcal{IG}(\alpha_f, \nu_f)$$
$$T_p \sim \mathcal{IG}(\alpha_p, \nu_p), \tag{10}$$

where we create some shorthands for convenience:

$$\alpha_f = \alpha(D_f), \nu_f = \nu(c_f, f_f, e_f = 0°)$$
$$\alpha_p = \alpha(D_p), \nu_p = \nu(c_p, f_p, e_p = 10°). \tag{11}$$

$e_p = 10°$ because the peripheral stimulus for this experiment was at $10°$ eccentricity. Then, as experimentally determined by prior literature on similar tasks [Ludwig et al. 2014], we model the total saccade latency as the maximum value of these two random variables:

$$T_{dual} = \max(T_f, T_p). \tag{12}$$

*Setup.* To evaluate our hypothetical model for dual tasks, we conducted a user study to measure how saccade latency changes as we modulate foveal and peripheral stimuli independently. Unfortunately, it is not possible to explicitly compute the $\alpha_f$ and $\alpha_p$ values as in Equation (7), because a user study for the dual task can only sample the *total* saccade latency from Equation (12). That is, the individual distributions, $T_f$, and $T_p$ are not measured directly. Since finding these threshold values directly is not possible, we infer them via maximum-likelihood estimation (MLE) of the overall distribution of $T_{dual}$, given a dataset of size $n$:

$$\alpha_f, \alpha_p = \arg\max \sum_i^n \log L(\alpha_f, \alpha_p; t^{(i)}, \nu_f, \nu_p). \tag{13}$$

Please refer to Appendix C for the derivation of the likelihood function for $T_{dual}$. The hardware setup in this experiment is the same as described in Table 1.

*Participants.* We recruited $n = 12$ participants (ages 22-33, 3 female) with normal or corrected to normal vision for a series of 2AFC experiments. The study was conducted during a single 10 minute session, including a total of 240 trials for each participant.

*Stimuli and Tasks.* At the beginning of each trial participants are shown three Gabor patches as illustrated in Figure 8a: one at the fovea, and two in the left and right peripheries at equal eccentricities of $10°$. The foveal Gabor is tilted either $45°$ or $-45°$ from the vertical axis; with chance probability, one of the peripheral Gabors is selected to have the same tilt as the foveal Gabor, while the other has the opposite tilt. The task is to identify and saccade to the peripheral Gabor of the same orientation as the foveal Gabor. For each trial, the central and peripheral Gabor contrast values are sampled from $[0.05, 0.22, 0.53, 1.0]$, drawn independently. That is, taking all combinations of central-peripheral Gabor contrast possibilities yields a total of 16 conditions. The frequency of all Gabors was fixed to 2.0 cpd (cycles-per-degree). Each participant also performed 15 randomly ordered practice trials before the start of the experiment.

*Results.* In Figure 8c, we show the relationship of both foveal and peripheral contrasts with saccade latency, as well as the ground truth data collected from our user study overlaid on top of the surface plot. The MLE regression produces threshold values of $\alpha_f = 3.21$ and $\alpha_p = 3.56$. Hence, the threshold ratio between the foveal and peripheral components is $1 : 1.04$. Similar to Section 6.1, we present the Q-Q plot comparing the data to our model predictions in Figure 8d. The K.S. statistical test again fails to reject the null hypothesis that the observed user saccadic latencies are drawn from our $T_{dual}$−predicted distribution ($D = 0.1$ and $p = 1.0$).

Models which consider only the peripheral contrast (shown in Figure 8b), or only the foveal contrast fail to accurately predict the saccade latencies. We run the K.S. test for both of these conditions and observe a significant difference between the data and the model predictions: $D = 0.9$ and $p = 0.002$ for the foveal-only model, and $D = 0.8$ and $p = 0.002$ for the periphery-only model.

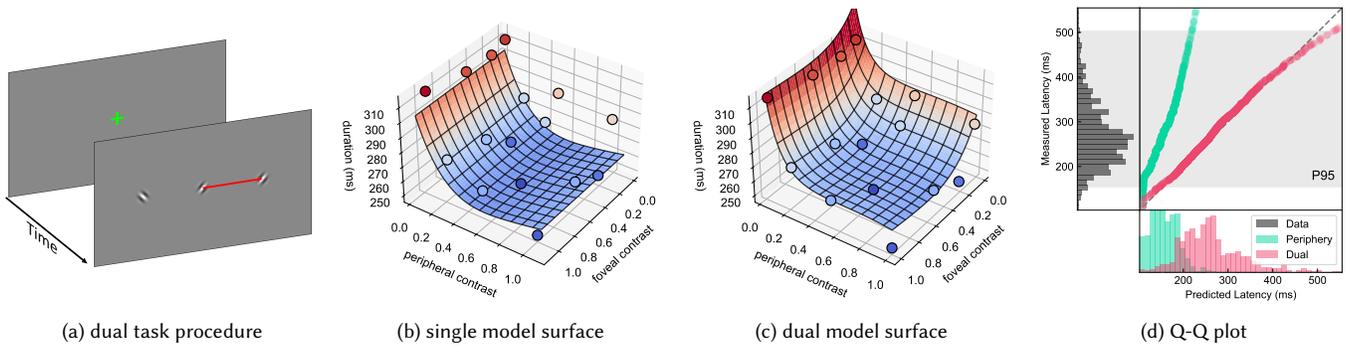

(a) dual task procedure   (b) single model surface   (c) dual model surface   (d) Q-Q plot

Fig. 8. *Model visualization and evaluation of foveal-peripheral dual task.* (a) A dual foveal-peripheral task consists of two components: identification of both the foveal and peripheral Gabor patches. The subject was instructed to move their gaze to the peripheral patch with matching orientation to the foveal one. Please refer to our supplementary videos for an animated visualization. (b) We fit our periphery-only model ($T_p$, the surface) to data from the foveal-peripheral dual task (the sparse dots). A significant mis-alignment can be observed. (c) Considering maximum expected latency of both foveal and peripheral contrasts enables us to predict a more accurate relationship, $T_{dual}$, between the visual stimulus parameters and the observed saccade latency data. (d) Q-Q plot visualizing the goodness-of-fit of our model relative to the observed data. Alignment of the observed and predicted latency histograms shows that the dual model matches well with the experimental data (gray) within the P95 confidence interval (highlighted region). In contrast, the peripheral-only model fails to correctly predict saccade latencies. We omit visualizations of the foveal-only model $T_f$ to avoid duplication as it exhibits similar low performance in predictive quality to $T_p$. The full statistical analysis can be seen in Section 6.3.

*Discussion.* When humans perform tasks involving both foveal and peripheral analysis, we observe that a models considering only one eccentricity fails to predict saccade latencies, as illustrated in Figure 8d and demonstrated by the K.S. tests. By comparison, the joint model we propose in Equation (12), inspired by prior discoveries on visual mechanisms, successfully predicts the latency distribution.

# 7 APPLICATION CASE STUDY: ESPORTS FAIRNESS METRIC AND PERFORMANCE OPTIMIZATION

A major application of our model is to measure and optimize human performance in competitive, real-time, or time-sensitive tasks such as defense, piloting, and esports. In this evaluation, we use esports as an example. In real-world professional gameplay, we deploy our model to: 1) measure game fairness in terms of character skin design that triggers varied gaze motion performance between two teams; 2) measure and optimize the human target search performance under various screen resolutions, eye-display distances, and compare the performance with traditional and immersive displays.

*Data.* We collected professional replay videos from a popular esports game, Counter-Strike: Global Offense via YouTube. The data contains a ≈ half hour long video footage where we uniformly sampled 95 frames from beginning to the end. For each frame, we exploit the virtual human tracking model YoLo [2016] that predicts the team ID (Counter-Terrorist, **CT** and Terrorist **TR**), and bounding boxes. We assume the observers gaze lies in the middle of the screen, and apply our model to predict the time when the viewer reorients their gaze to each target. We measure the visual characteristics with a common display setting: a Samsung 32inch CH32H711 monitor, 2K 16:9 resolution, 70cm width, 300cd/$m^2$ brightness, and ≈ 1.33D (diopter = $m^{-1}$) eye-display distance (50° FoV).

## 7.1 Competition Fairness in Target Searching

The game has two opposing teams of characters. Regardless of game task design and differences in tools, game fairness is an important concern in esports [Chen et al. 2014]. Using our model and the detected targets, we measure the average saccade latency of individual groups.

Figure 9b shows the results. We observed a significant difference between **CT** and **TR** groups: the average normalized latencies are $0.92 \pm 0.02$ for searching **CT**s and $0.95 \pm 0.04$ for searching **TR**s, indicating a 3.3% difference. Given previous literature indicating the mean saccade latency for CS:GO professional players to be about 282ms [Velichkovsky et al. 2019], 3.3% results in a 9.3ms reaction variance. One-way repeated measures ANOVA showed the group's significant main effect on the saccadic latency, $F_{1,93} = 11.4, p = .001$.

The results demonstrate a statistically significant difference between the two groups, in terms of them perceiving, processing, and reacting to appeared targets. That is, a **TR/CT** saccading to the other group is significantly faster/slower with no less than 2/1 frames on a 60/120FPS displays. The speed difference is remarkably higher than the minimum latency, as low as 4ms, that leads to altered esport performance among top-level competitors [Kim et al. 2019]. While this may have been one of the factors that contributed to the imbalanced competitive game performance (higher winning rate of **TR** on the map we analyzed) between these two groups [1], in practice the asymmetric weapon and task designs might also have played a role.

## 7.2 Optimizing Player Performance

A natural and extensively asked question is the role of eye-display distance (e.g., regular monitors vs. VR displays) and screen resolution in professional competitions. Using our model, we measure the

---

statistical saccade latency as a function of displays with the same dataset as Section 7.

Figure 9c visualizes our results by observing the altered reaction performance. As before, we use the mean saccade latency for CS:GO professional players to be about 282ms [Velichkovsky et al. 2019]. First, both teams are not at their best performance with the initial 1.33D eye-display distance, with the faster reaction of **TR**s searching **CT**s. However, the teams reveal different trends by changing the eye's distance (thus FoV), which jointly alters target eccentricity (*e*) and frequency (*f*) (cf. Appendix D). Particularly, the minimal saccade latency towards **CT** targets is 273.5ms at 34.6° FoV (0.9D eye-display distance). In comparison, the minimal latency towards **TR** targets is 266.1ms at 61.5° FoV (1.6D eye-display distance). The two curves intersect at 1.69D with an identical latency of 266.3ms. We further simulate real-world use cases with different displays (for instance, gaming with mobile devices or training with VR displays). In this experiment, we use the measures from the iPhone 13 (5.78 × 2.53 inches) with the commonly suggested 30cm (or 3.3D) eye-display distance, leading to a 25.7° FoV. Under this circumstances, the saccade time to **TR** becomes higher than **CT** (324.3ms vs 259.4ms) Similarly, the measurement with our virtual reality HMD (90° overlapped FoV), the relative trend is swapped: saccading to a **TR** becomes shorter than to a **CT**, with a 21.2ms difference (276.0ms vs 288.0ms).

The above analysis indicates the sensitivity of eye display correlation in determining performance and fairness in time-sensitive and competitive scenarios. Surprisingly, the statistical performance bias may swap with different eye-display relationships. For instance, with a mobile/VR setting, the visual stimuli may bias with **TR/CT** players in terms of reaction performance. In addition to the commonly referred measurement of visual similarity and task/map fairness, our model presents a novel perspective in competitive and highly dynamic scenario design, such as athletics, esports, and defense.

## 8 LIMITATIONS AND FUTURE WORK

In our model, we consider a variety of factors which affect visual stimuli – contrast, frequency, eccentricity – in the context of several display environments (from mobile devices to VR displays). However, a multitude of other factors may affect reaction time to observed natural stimuli in the complex natural world. For instance, making a cognitive decision among multiple potential targets may depend on higher-level visual salience [Jarvenpaa 1990] and object sizes [Lisi et al. 2019] whereas our model assumes pre-knowledge of the intended saccade target. While our work showed promising results for one foveal and one peripheral task, we cannot easily extrapolate our results for significantly more complex tasks. In addition, we rely on a single max-pooling of image frequencies using the highest contrast layer of the Laplacian pyramid (Section 6.2) for model application in natural images. This may introduce an approximation error, especially with complex and noisy natural images (Figure 7c). Using an approach similar to Mantiuk et al. [2021] and fitting a pooling model to user data collected from natural image stimuli would be a useful direction of future work. To avoid any bias introduced by color, we convert stimuli to gray-scale for the psychophysical experiments. We leave the task of creating a multi-band and color-aware model tailored for measuring reactive latency for future work. Our model also does not consider scenarios where motion [Jindal et al. 2021] and refresh rates [Krajancich et al. 2021] may play a critical role, especially during saccades [Schweitzer and Rolfs 2021]. We aim to extend our model to consider spatio-temporal effects and complex dynamic scenarios in the future.

## 9 CONCLUSION

We demonstrated a significant difference between human visual acuity (*observation* of a stimulus) and reaction latency (processing *after* the observation). We formulated this behavior using a neurologically-inspired probabilistic model that is motivated and evaluated by a series of psychophysical studies. The surprising gap we observe between observation and reaction raised new, previously unasked questions such as, "Are competitive digital activities such as esports fair among teams, given the design of their appearances?", "How do human-display-content relationships alter our performance in virtual environments?", and "What settings optimize reaction time without compromising visual content?". We hope that our model's answers to these questions inspire researchers to explore new avenues in interactive computer graphics and immersive virtual environments.

## ACKNOWLEDGMENTS

The authors would like to thank Chris Wyman for his valuable suggestions. The research is partially supported by the NVIDIA Applied Research Accelerator Program and the DARPA PTG program. Any opinions, findings, and conclusions or recommendations expressed in this material are those of the authors and do not necessarily reflect the views of DARPA.

## REFERENCES

Rachel Albert, Anjul Patney, David Luebke, and Joohwan Kim. 2017. Latency Requirements for Foveated Rendering in Virtual Reality. *ACM Transactions on Applied Perception* 14, 4, Article 25 (sep 2017), 13 pages. https://doi.org/10.1145/3127589

Elena Arabadzhiyska, Okan Tarhan Tursun, Karol Myszkowski, Hans-Peter Seidel, and Piotr Didyk. 2017. Saccade Landing Position Prediction for Gaze-Contingent Rendering. *ACM Trans. Graph.* 36, 4, Article 50 (July 2017), 12 pages. https://doi.org/10.1145/3072959.3073642

A.Terry Bahill, Michael R. Clark, and Lawrence Stark. 1975. The main sequence, a tool for studying human eye movements. *Mathematical Biosciences* 24, 3 (1975), 191–204. https://doi.org/10.1016/0025-5564(75)90075-9

A Terry Bahill. 1975. Most naturally occurring human saccades have magnitudes of 15 deg or less. *Invest. Ophthalmol* 14 (1975), 468–469.

Reynold Bailey, Ann McNamara, Nisha Sudarsanam, and Cindy Grimm. 2009. Subtle Gaze Direction. *ACM Trans. Graph.* 28, 4, Article 100 (Sept. 2009), 14 pages. https://doi.org/10.1145/1559755.1559757

Peter GJ Barten. 1999. *Contrast sensitivity of the human eye and its effects on image quality.* SPIE press.

W. Becker and A.F. Fuchs. 1969. Further properties of the human saccadic system: Eye movements and correction saccades with and without visual fixation points. *Vision Research* 9, 10 (1969), 1247–1258. https://doi.org/10.1016/0042-6989(69)90112-6

AH Bell, MA Meredith, AJ Van Opstal, and DougP Munoz. 2006. Stimulus intensity modifies saccadic reaction time and visual response latency in the superior colliculus. *Experimental Brain Research* 174, 1 (2006), 53–59.

DC Burr, MC Morrone, and J Ross. 1994. Selective suppression of the magnocellular visual pathway during saccadic eye movements. *Nature* 371, 6497 (1994), 511–513.

Anke Cajar, Ralf Engbert, and Jochen Laubrock. 2016. Spatial frequency processing in the central and peripheral visual field during scene viewing. *Vision Research* 127 (2016), 186–197.

RHS Carpenter. 2004. Contrast, probability, and saccadic latency: evidence for independence of detection and decision. *Current Biology* 14, 17 (2004), 1576–1580.

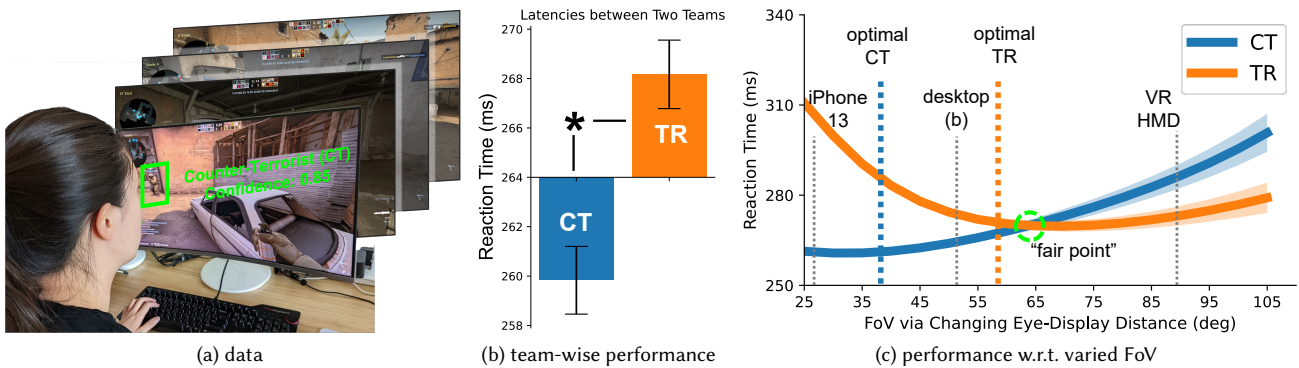

(a) data    (b) team-wise performance    (c) performance w.r.t. varied FoV

Fig. 9. *Results of esports video dataset analysis* (a) illustrates our simulated eye-display spatial relationship and our CS:GO gameplay dataset (including the automated labeling of the teams). Note that changing the eye-display distances results in varied FoVs, thus changing the perceived visual characteristics (eccentricity and frequency). (b) shows our model's approximation of the team-wise target searching performance. The X-axis indicates the team splits. The Y-axis shows the mean saccadic latency calculated by our model (with the annotated team as the target team being searched, i.e., the "opposite team"). The error bars show the standard error. (c) shows our analysis simulating various FoVs by altering eye-display distances. The X-axis indicates the FoV in degrees. The Y-axis shows the predicted mean latencies with the semi-transparent error bar as standard error. The point where the two group's mean latencies intersect is marked by the green circle. The lowest latencies of saccading for each team and the simulated FoVs of non-desktop display environments are dash-labeled.

Roger HS Carpenter and MLL Williams. 1995. Neural computation of log likelihood in control of saccadic eye movements. *Nature* 377, 6544 (1995), 59–62.

Haoyang Chen, Yasukuni Mori, and Ikuo Matsuba. 2014. Solving the balance problem of massively multiplayer online role-playing games using coevolutionary programming. *Applied Soft Computing* 18 (2014), 1–11.

Shaoyu Chen, Budmonde Duinkharjav, Xin Sun, Li-Yi Wei, Stefano Petrangeli, Jose Echevarria, Claudio Silva, and Qi Sun. 2022. Instant Reality: Gaze-Contingent Perceptual Optimization for 3D Virtual Reality Streaming. *IEEE Transactions on Visualization and Computer Graphics* 28, 5 (2022), 2157–2167. https://doi.org/10.1109/TVCG.2022.3150522

Michael A. Cohen, Thomas L. Botch, and Caroline E. Robertson. 2020. The limits of color awareness during active, real-world vision. *Proceedings of the National Academy of Sciences* 117, 24 (2020), 13821–13827. https://doi.org/10.1073/pnas.1922294117 arXiv:https://www.pnas.org/content/117/24/13821.full.pdf

Julien Cotti, Muriel Panouilleres, Douglas P Munoz, Jean-Louis Vercher, Denis Pélisson, and Alain Guillaume. 2009. Adaptation of reactive and voluntary saccades: different patterns of adaptation revealed in the antisaccade task. *The Journal of Physiology* 587, 1 (2009), 127–138.

Scott J Daly. 1992. Visible differences predictor: an algorithm for the assessment of image fidelity. In *Human Vision, Visual Processing, and Digital Display III*, Vol. 1666. International Society for Optics and Photonics, 2–15.

H. Deubel, W. Wolf, and G. Hauske. 1982. Corrective saccades: Effect of shifting the saccade goal. *Vision Research* 22, 3 (1982), 353–364. https://doi.org/10.1016/0042-6989(82)90151-1

Mark R. Diamond, John Ross, and M. C. Morrone. 2000. Extraretinal Control of Saccadic Suppression. *Journal of Neuroscience* 20, 9 (2000), 3449–3455. https://doi.org/10.1523/JNEUROSCI.20-09-03449.2000 arXiv:https://www.jneurosci.org/content/20/9/3449.full.pdf

Andrew T. Duchowski, Donald H. House, Jordan Gestring, Rui I. Wang, Krzysztof Krejtz, Izabela Krejtz, Radosław Mantiuk, and Bartosz Bazyluk. 2014. Reducing Visual Discomfort of 3D Stereoscopic Displays with Gaze-Contingent Depth-of-Field (*SAP '14*). Association for Computing Machinery, New York, NY, USA, 39–46. https://doi.org/10.1145/2628257.2628259

David Dunn, Okan Tursun, Hyeonseung Yu, Piotr Didyk, Karol Myszkowski, and Henry Fuchs. 2020. Stimulating the Human Visual System Beyond Real World Performance in Future Augmented Reality Displays. In *2020 IEEE International Symposium on Mixed and Augmented Reality (ISMAR)*. IEEE, 90–100.

Ralf Engbert and Konstantin Mergenthaler. 2006. Microsaccades are triggered by low retinal image slip. *Proceedings of the National Academy of Sciences* 103, 18 (2006), 7192–7197.

Jasper H Fabius, Alessio Fracasso, Tanja CW Nijboer, and Stefan Van der Stigchel. 2019. Time course of spatiotopic updating across saccades. *Proceedings of the National Academy of Sciences* 116, 6 (2019), 2027–2032.

J Leroy Folks and Raj S Chhikara. 1978. The inverse Gaussian distribution and its statistical application—a review. *Journal of the Royal Statistical Society: Series B (Methodological)* 40, 3 (1978), 263–275.

Linus Franke, Laura Fink, Jana Martschinke, Kai Selgrad, and Marc Stamminger. 2021. Time-Warped Foveated Rendering for Virtual Reality Headsets. *Computer Graphics Forum* 40, 1 (2021), 110–123. https://doi.org/10.1111/cgf.14176 arXiv:https://onlinelibrary.wiley.com/doi/pdf/10.1111/cgf.14176

Drew Fudenberg, Whitney Newey, Philipp Strack, and Tomasz Strzalecki. 2020. Testing the drift-diffusion model. *Proceedings of the National Academy of Sciences* 117, 52 (2020), 33141–33148.

Ramanathan Gnanadesikan and Martin B Wilk. 1968. Probability plotting methods for the analysis of data. *Biometrika* 55, 1 (1968), 1–17.

Brian Guenter, Mark Finch, Steven Drucker, Desney Tan, and John Snyder. 2012. Foveated 3D Graphics. *ACM Transactions on Graphics* 31, 6, Article 164 (nov 2012), 10 pages. https://doi.org/10.1145/2366145.2366183

E. Hartmann, B. Lachenmayr, and H. Brettel. 1979. The peripheral critical flicker frequency. *Vision Research* 19, 9 (1979), 1019–1023. https://doi.org/10.1016/0042-6989(79)90227-X

Toyohiko Hatada, Haruo Sakata, and Hideo Kusaka. 1980. Psychophysical analysis of the "sensation of reality" induced by a visual wide-field display. *Smpte Journal* 89, 8 (1980), 560–569.

Sebastien Hillaire, Anatole Lecuyer, Remi Cozot, and Gery Casiez. 2008. Using an Eye-Tracking System to Improve Camera Motions and Depth-of-Field Blur Effects in Virtual Environments. In *2008 IEEE Virtual Reality Conference*. 47–50. https://doi.org/10.1109/VR.2008.4480749

Alain Hore and Djemel Ziou. 2010. Image quality metrics: PSNR vs. SSIM. In *2010 20th international conference on pattern recognition*. IEEE, 2366–2369.

Michael R. Ibbotson and Shaun L. Cloherty. 2009. Visual Perception: Saccadic Omission—Suppression or Temporal Masking? *Current Biology* 19, 12 (2009), R493–R496. https://doi.org/10.1016/j.cub.2009.05.010

Sirkka L Jarvenpaa. 1990. Graphic displays in decision making—the visual salience effect. *Journal of Behavioral Decision Making* 3, 4 (1990), 247–262.

Akshay Jindal, Krzysztof Wolski, Karol Myszkowski, and Rafal K Mantiuk. 2021. Perceptual model for adaptive local shading and refresh rate. *ACM Transactions on Graphics (TOG)* 40, 6 (2021), 1–18.

RP Kalesnykas and PE Hallett. 1994. Retinal eccentricity and the latency of eye saccades. *Vision research* 34, 4 (1994), 517–531.

Anton S Kaplanyan, Anton Sochenov, Thomas Leimkühler, Mikhail Okunev, Todd Goodall, and Gizem Rufo. 2019. DeepFovea: neural reconstruction for foveated rendering and video compression using learned statistics of natural videos. *ACM Transactions on Graphics (TOG)* 38, 6 (2019), 1–13.

Donald H Kelly. 1979. Motion and vision. II. Stabilized spatio-temporal threshold surface. *Josa* 69, 10 (1979), 1340–1349.

Joohwan Kim, Josef Spjut, Morgan McGuire, Alexander Majercik, Ben Boudaoud, Rachel Albert, and David Luebke. 2019. Esports arms race: Latency and refresh rate for competitive gaming tasks. *Journal of Vision* 19, 10 (2019), 218c–218c.

Robert Konrad, Anastasios Angelopoulos, and Gordon Wetzstein. 2020. Gaze-Contingent Ocular Parallax Rendering for Virtual Reality. *ACM Trans. Graph.* 39 (2020). Issue 2.

Denis Koposov, Maria Semenova, Andrey Somov, Andrey Lange, Anton Stepanov, and Evgeny Burnaev. 2020. Analysis of the reaction time of esports players through the gaze tracking and personality trait. In *2020 IEEE 29th International Symposium on Industrial Electronics (ISIE)*. IEEE, 1560–1565.

Matias Koskela, Atro Lotvonen, Markku Mäkitalo, Petrus Kivi, Timo Viitanen, and Pekka Jääskeläinen. 2019. Foveated real-time path tracing in visual-polar space. In *Eurographics Symposium on Rendering*. The Eurographics Association.

Matias Koskela, Timo Viitanen, Pekka Jääskeläinen, and Jarmo Takala. 2016. Foveated path tracing. In *International Symposium on Visual Computing*. Springer, 723–732.

Eileen Kowler. 2011. Eye movements: The past 25years. *Vision Research* 51, 13 (2011), 1457–1483. https://doi.org/10.1016/j.visres.2010.12.014 Vision Research 50th Anniversary Issue: Part 2.

Brooke Krajancich, Petr Kellnhofer, and Gordon Wetzstein. 2020. Optimizing depth perception in virtual and augmented reality through gaze-contingent stereo rendering. *ACM Transactions on Graphics (TOG)* 39, 6 (2020), 1–10.

Brooke Krajancich, Petr Kellnhofer, and Gordon Wetzstein. 2021. A Perceptual Model for Eccentricity-dependent Spatio-temporal Flicker Fusion and its Applications to Foveated Graphics. *ACM Trans. Graph.* 40 (2021). Issue 4.

Matteo Lisi, Joshua A. Solomon, and Michael J. Morgan. 2019. Gain control of saccadic eye movements is probabilistic. *Proceedings of the National Academy of Sciences* 116, 32 (2019), 16137–16142. https://doi.org/10.1073/pnas.1901963116 arXiv:https://www.pnas.org/content/116/32/16137.full.pdf

Casimir JH Ludwig, J Rhys Davies, and Miguel P Eckstein. 2014. Foveal analysis and peripheral selection during active visual sampling. *Proceedings of the National Academy of Sciences* 111, 2 (2014), E291–E299.

Madhumitha S Mahadevan, Harold E Bedell, and Scott B Stevenson. 2018. The influence of endogenous attention on contrast perception, contrast discrimination, and saccadic reaction time. *Vision research* 143 (2018), 89–102.

Rafal Mantiuk, Grzegorz Krawczyk, Karol Myszkowski, and Hans-Peter Seidel. 2004. Perception-motivated high dynamic range video encoding. *ACM Transactions on Graphics (TOG)* 23, 3 (2004), 733–741.

Rafał K Mantiuk, Gyorgy Denes, Alexandre Chapiro, Anton Kaplanyan, Gizem Rufo, Romain Bachy, Trisha Lian, and Anjul Patney. 2021. FovVideoVDP: A visible difference predictor for wide field-of-view video. *ACM Transactions on Graphics (TOG)* 40, 4 (2021), 1–19.

Frank J Massey Jr. 1951. The Kolmogorov-Smirnov test for goodness of fit. *Journal of the American statistical Association* 46, 253 (1951), 68–78.

Ethel Matin. 1975. Saccadic suppression: A review and an analysis. *Psychological bulletin* 81 (01 1975), 899–917. https://doi.org/10.1037/h0037368

Michael Mauderer, Simone Conte, Miguel A. Nacenta, and Dhanraj Vishwanath. 2014. Depth Perception with Gaze-Contingent Depth of Field. In *Proceedings of the SIGCHI Conference on Human Factors in Computing Systems (CHI '14)*. Association for Computing Machinery, New York, NY, USA, 217–226. https://doi.org/10.1145/2556288.2557089

Deepmala Mazumdar, Najiya S Kadavath Meethal, Manish Panday, Rashima Asokan, Gijs Thepass, Ronnie J George, Johannes van der Steen, and Johan JM Pel. 2019. Effect of age, sex, stimulus intensity, and eccentricity on saccadic reaction time in eye movement perimetry. *Translational Vision Science & Technology* 8, 4 (2019), 13–13.

Mark E Mazurek, Jamie D Roitman, Jochen Ditterich, and Michael N Shadlen. 2003. A role for neural integrators in perceptual decision making. *Cerebral cortex* 13, 11 (2003), 1257–1269.

Suzanne P McKee and Ken Nakayama. 1984. The detection of motion in the peripheral visual field. *Vision research* 24, 1 (1984), 25–32.

Xiaoxu Meng, Ruofei Du, Matthias Zwicker, and Amitabh Varshney. 2018. Kernel Foveated Rendering. *Proceedings of ACM Computer Graphics and Interactive Techniques* 1, 1, Article 5 (jul 2018), 20 pages. https://doi.org/10.1145/3203199

Aythami Morales, Francisco M Costela, and Russell L Woods. 2021. Saccade Landing Point Prediction Based on Fine-Grained Learning Method. *IEEE Access* 9 (2021), 52474–52484.

Manon Mulckhuyse and Jan Theeuwes. 2010. Unconscious cueing effects in saccadic eye movements–Facilitation and inhibition in temporal and nasal hemifield. *Vision Research* 50, 6 (2010), 606–613.

Cornelis Noorlander, Jan J. Koenderink, Ron J. Den Olden, and B. Wigbold Edens. 1983. Sensitivity to spatiotemporal colour contrast in the peripheral visual field. *Vision Research* 23, 1 (1983), 1–11.

Evan M Palmer, Todd S Horowitz, Antonio Torralba, and Jeremy M Wolfe. 2011. What are the shapes of response time distributions in visual search? *Journal of experimental psychology: human perception and performance* 37, 1 (2011), 58.

John Palmer, Alexander C Huk, and Michael N Shadlen. 2005. The effect of stimulus strength on the speed and accuracy of a perceptual decision. *Journal of vision* 5, 5 (2005), 1–1.

Anjul Patney, Marco Salvi, Joohwan Kim, Anton Kaplanyan, Chris Wyman, Nir Benty, David Luebke, and Aaron Lefohn. 2016. Towards Foveated Rendering for Gaze-Tracked Virtual Reality. *ACM Trans. Graph.* 35, 6, Article 179 (Nov. 2016), 12 pages. https://doi.org/10.1145/2980179.2980246

Andreas Polychronakis, George Alex Koulieris, and Katerina Mania. 2021. Emulating Foveated Path Tracing. In *Motion, Interaction and Games*. 1–9.

Dale Purves, Roberto Cabeza, Scott A Huettel, Kevin S LaBar, Michael L Platt, Marty G Woldorff, and Elizabeth M Brannon. 2008. *Cognitive neuroscience*. Sunderland: Sinauer Associates, Inc.

Roger Ratcliff. 1978. A theory of memory retrieval. *Psychological review* 85, 2 (1978), 59.

Baj AJ Reddi, Kaleab N Asrress, and Roger HS Carpenter. 2003. Accuracy, information, and response time in a saccadic decision task. *Journal of neurophysiology* 90, 5 (2003), 3538–3546.

Joseph Redmon, Santosh Divvala, Ross Girshick, and Ali Farhadi. 2016. You only look once: Unified, real-time object detection. In *Proceedings of the IEEE conference on computer vision and pattern recognition*. 779–788.

Snježana Rimac-Drlje, Mario Vranješ, and Drago Žagar. 2010. Foveated Mean Squared Error–a Novel Video Quality Metric. *Multimedia Tools and Applications* 49, 3 (sep 2010), 425–445. https://doi.org/10.1007/s11042-009-0442-1

Snježana Rimac-Drlje, Goran Martinović, and Branka Zovko-Cihlar. 2011. Foveation-based content Adaptive Structural Similarity index. In *2011 18th International Conference on Systems, Signals and Image Processing*. 1–4.

Richard Schweitzer and Martin Rolfs. 2021. Intrasaccadic motion streaks jump-start gaze correction. *Science Advances* 7, 30 (2021), eabf2218.

Ana Serrano, Vincent Sitzmann, Jaime Ruiz-Borau, Gordon Wetzstein, Diego Gutierrez, and Belen Masia. 2017. Movie editing and cognitive event segmentation in virtual reality video. *ACM Transactions on Graphics (TOG)* 36, 4 (2017), 1–12.

Vincent Sitzmann, Ana Serrano, Amy Pavel, Maneesh Agrawala, Diego Gutierrez, Belen Masia, and Gordon Wetzstein. 2018. Saliency in VR: How do people explore virtual environments? *IEEE transactions on visualization and computer graphics* 24, 4 (2018), 1633–1642.

Miriam Spering and Marisa Carrasco. 2015. Acting without seeing: eye movements reveal visual processing without awareness. *Trends in neurosciences* 38, 4 (2015), 247–258.

Qi Sun, Fu-Chung Huang, Joohwan Kim, Li-Yi Wei, David Luebke, and Arie Kaufman. 2017. Perceptually-Guided Foveation for Light Field Displays. *ACM Trans. Graph.* 36, 6, Article 192 (Nov. 2017), 13 pages. https://doi.org/10.1145/3130800.3130807

Qi Sun, Fu-Chung Huang, Li-Yi Wei, David Luebke, Arie Kaufman, and Joohwan Kim. 2020. Eccentricity effects on blur and depth perception. *Optics express* 28, 5 (2020), 6734–6739.

Qi Sun, Anjul Patney, Li-Yi Wei, Omer Shapira, Jingwan Lu, Paul Asente, Suwen Zhu, Morgan Mcguire, David Luebke, and Arie Kaufman. 2018. Towards Virtual Reality Infinite Walking: Dynamic Saccadic Redirection. *ACM Trans. Graph.* 37, 4, Article 67 (July 2018), 13 pages. https://doi.org/10.1145/3197517.3201294

The Manim Community Developers. 2022. *Manim – Mathematical Animation Framework*. https://www.manim.community/

L.N. Thibos, D.J. Walsh, and F.E. Cheney. 1987b. Vision beyond the resolution limit: Aliasing in the periphery. *Vision Research* 27, 12 (1987), 2193–2197.

L. N. Thibos, F. E. Cheney, and D. J. Walsh. 1987a. Retinal limits to the detection and resolution of gratings. *Journal of the Optical Society of America A* 4, 8 (1987), 1524–1529.

Okan Tarhan Tursun, Elena Arabadzhiyska-Koleva, Marek Wernikowski, Radosław Mantiuk, Hans-Peter Seidel, Karol Myszkowski, and Piotr Didyk. 2019. Luminance-contrast-aware foveated rendering. *ACM Transactions on Graphics (TOG)* 38, 4 (2019), 1–14.

Robert J van Beers. 2007. The sources of variability in saccadic eye movements. *Journal of Neuroscience* 27, 33 (2007), 8757–8770.

Boris B Velichkovsky, Nikita Khromov, Alexander Korotin, Evgeny Burnaev, and Andrey Somov. 2019. Visual fixations duration as an indicator of skill level in esports. In *IFIP Conference on Human-Computer Interaction*. Springer, 397–405.

David R Walton, Rafael Kuffner Dos Anjos, Sebastian Friston, David Swapp, Kaan Akşit, Anthony Steed, and Tobias Ritschel. 2021. Beyond blur: Real-time ventral metamers for foveated rendering. *ACM Transactions on Graphics* 40, 4 (2021), 1–14.

Zhou Wang, Alan Conrad Bovik, Ligang Lu, and Jack L. Kouloheris. 2001. Foveated wavelet image quality index. In *Applications of Digital Image Processing XXIV*, Andrew G. Tescher (Ed.), Vol. 4472. International Society for Optics and Photonics, SPIE, 42 – 52. https://doi.org/10.1117/12.449797

Martin Weier, Thorsten Roth, Ernst Kruijff, André Hinkenjann, Arsène Pérard-Gayot, Philipp Slusallek, and Yongmin Li. 2016. Foveated real-time ray tracing for head-mounted displays. In *Computer Graphics Forum*, Vol. 35. Wiley Online Library, 289–298.

Shimpei Yamagishi and Shigeto Furukawa. 2020. Factors Influencing Saccadic Reaction Time: Effect of Task Modality, Stimulus Saliency, Spatial Congruency of Stimuli, and Pupil Size. *Frontiers in Human Neuroscience* (2020), 513.

## A  DERIVING Equation (5)

For a Brownian motion process as described by Equation (1),

$$A(0; v) = 0,$$
$$A(t; v) = vt + W(t)$$
$$W(t) \sim \mathcal{G}(0, t),$$

the joint probability distribution of an evidence value $a$ observed at time $t$ is described by the Fokker-Plank equation:

$$\frac{\partial p}{\partial t} + v \frac{\partial p}{\partial a} = \frac{1}{2} \frac{\partial^2 p}{\partial^2 a}, \qquad (14)$$

with boundary conditions

$$\begin{cases} p(0, a) &= \delta(a) \\ p(t, \alpha) &= 0 \end{cases} \qquad (15)$$

where $p$ is the probability density function of particles behaving according to Equation (1), and $\delta$ is the Dirac delta function. The solution to the boundary value problem described by Equation (14), with boundary conditions of Equation (15), is

$$p(t, a) = \frac{1}{\sqrt{2\pi t}} \left( \exp\left[ -\frac{(a - vt)^2}{2t} \right] - \exp\left[ 2v\alpha - \frac{(a - 2\alpha - vt)^2}{2t} \right] \right). \qquad (16)$$

This probability density function describes the joint probability of observing any given pair of time $t$ and evidence $a$. Using this density function, we first compute the probability of the evidence being below the threshold, $\alpha$. For the distribution of first passage time, $T$, this probability is equivalent to the survival function. I.e.,

$$S(t) = P(T > t) = \int_{-\infty}^{\alpha} p(t, a) da. \qquad (17)$$

Plugging in Equation (16) into Equation (17) we get,

$$S(t) = \Phi\left( \frac{\alpha - vt}{\sqrt{t}} \right) - \exp(2v\alpha)\Phi\left( \frac{-\alpha - vt}{\sqrt{t}} \right). \qquad (18)$$

Finally, we are able to derive the probability density function of $T$ via the relation between the PDF function and the survival function:

$$h(t) = -\frac{dS}{dt}$$
$$= \frac{\alpha}{\sqrt{2\pi t^3}} \exp \frac{-(\alpha - vt)^2}{2t}. \qquad (19)$$

## B  DATA NORMALIZATION PSEUDO-CODE

We describe the normalization and calibration procedures applied that are necessary for optimizing and subsequently using the model for novel applications.

**Require:** pilot study data
1: pick a pedestal condition (e.g. $c = 1.0$, $f = 1.0$, $e = 0°$)
2: **for** each subject **do**
3:    compute the average latency of pedestal condition, $t_{pedestal}$
4:    scale **all** latencies by $1/t_{pedestal}$
5: **end for**
6: train RBF network for computing $v$ in normalized units
7: **return**  normalized $v$ predictor.

Algorithm 1.  Normalization.

Once we obtain an optimized $v$ predictor, we apply the model to a novel application as follows:

**Require:** target application sample data
1: measure the $\mathbb{E}$ and $\mathbb{V}$ of the latency
2: compute $\alpha$ for data using eq. (7)
3: rescale $v$ by $\mathbb{E}$ of the data
4: **return**  probability distribution described by $\alpha$ and rescaled $v$.

Algorithm 2.  Calibration.

Due to the inverse correlation between the step 4 in Alg. 1 and the step 3 in Alg. 2, any selection of condition in step 1 of Alg. 1 does not lose the generality.

## C  DERIVING Equation (13)

We are interested in deriving an expression for the probability distribution function for $T_{dual}$ as shown in Equation (12).

$$T_{dual} = \max(T_f, T_p).$$

We know that both $T_f$ and $T_p$ are Inverse Gaussian (IG) random variables as detailed in Equation (10),

$$T_f \sim \mathcal{IG}(\alpha_f, v_f)$$
$$T_p \sim \mathcal{IG}(\alpha_p, v_p).$$

The probability that $T_{dual}$ is less than some time $t$ is equivalent to the statement that both $T_f$ and $T_p$ are less than $t$. I.e.,

$$P(T_{dual} \leq t) = P(T_f \leq t)P(T_p \leq t), \qquad (20)$$

or,

$$H_{dual}(t) = H_f(t)H_p(t), \qquad (21)$$

where $H_f$ denotes the cumulative density function (CDF) of the IG distribution with parameters $\alpha_f$ and $v_f$, and vice versa for $H_p$. The probability density function of $T_{dual}$ is therefore equal to the derivative of $H_{dual}$.

Taking the derivative from Equation (21) we get,

$$h_{dual}(t) = h_f(t)H_p(t) + H_f(t)h_p(t). \qquad (22)$$

Since we have an explicit expression for the PDF of $T_{dual}$, we can finally write down an expression for the likelihood function from Equation (13) as

$$L(\alpha_f, \alpha_p; t, v_f, v_p) = h(t; \alpha_f, v_f)H(t; \alpha_p, v_p) + \\ + H(t; \alpha_f, v_f)h(t; \alpha_p, v_p), \qquad (23)$$

where $h$ and $H$ are the PDF, and CDF functions of the IG distribution.

## D  FIELD-OF-VIEW VS ECCENTRICITY & FREQUENCY

The observed image characteristics of stimuli shown on a display vary depending on how far the display is from the eye. We correlate these effects using the field-of-view that the display occupies as a measure of eye-distance. FoV is an intuitive way to measure eye-distance as it can be used regardless of the specific dimensions of a given display.

Given a display with width $w$, presented at an FoV of $\theta_{\text{fov}}$, the distance of the display equals

$$d = \frac{w/2}{\tan(\theta_{\text{fov}}/2)}. \tag{24}$$

If an observer is staring at the center of the display at FoV of $\theta_{\text{fov}}$ (or equivalently at a distance of $d$), an object $x$cm away from the center of the display will appear at

$$\theta = \arctan \frac{x}{d} = \arctan\left(x\frac{\tan(\theta_{\text{fov}}/2)}{w/2}\right) \tag{25}$$

retinal eccentricity. Hence, we notice that changing the eye-distance of a display alters the eccentricity at which stimuli appear in the retina.

Additionally, we can use this relation to derive a rate-of-change coefficient between physical distances (in cm), and retinal eccentricities (in degrees) by taking the derivative of eq. (25),

$$\frac{d\theta}{dx} = \frac{\cos^2\theta}{d} = \cos^2\theta \frac{\tan(\theta_{\text{fov}}/2)}{w/2}. \tag{26}$$

This measure of "degrees-per-distance" allows us to derive the relationship between the spatial frequency of a pattern shown on the screen, $f_{\text{display}}$ (in cycles-per-centimeter), and the retinal frequency that an observer perceives, $f_{\text{retina}}$ (in cycles-per-degrees),

$$f_{\text{retina}} = f_{\text{display}} \frac{1}{\cos^2\theta} \frac{w/2}{\tan(\theta_{\text{fov}}/2)}. \tag{27}$$

Note that the observed frequency not only depends on the FoV, but also the eccentricity at which the stimulus is shown. For the simplest case where the stimulus is at the center of the screen, or $\theta = 0$, the relationship simplifies to

$$f_{\text{retina}} = f_{\text{display}} \frac{w/2}{\tan(\theta_{\text{fov}}/2)}. \tag{28}$$

## E PLOTS FOR INDIVIDUAL PARTICIPANTS IN PRELIMINARY STUDY (Section 3)

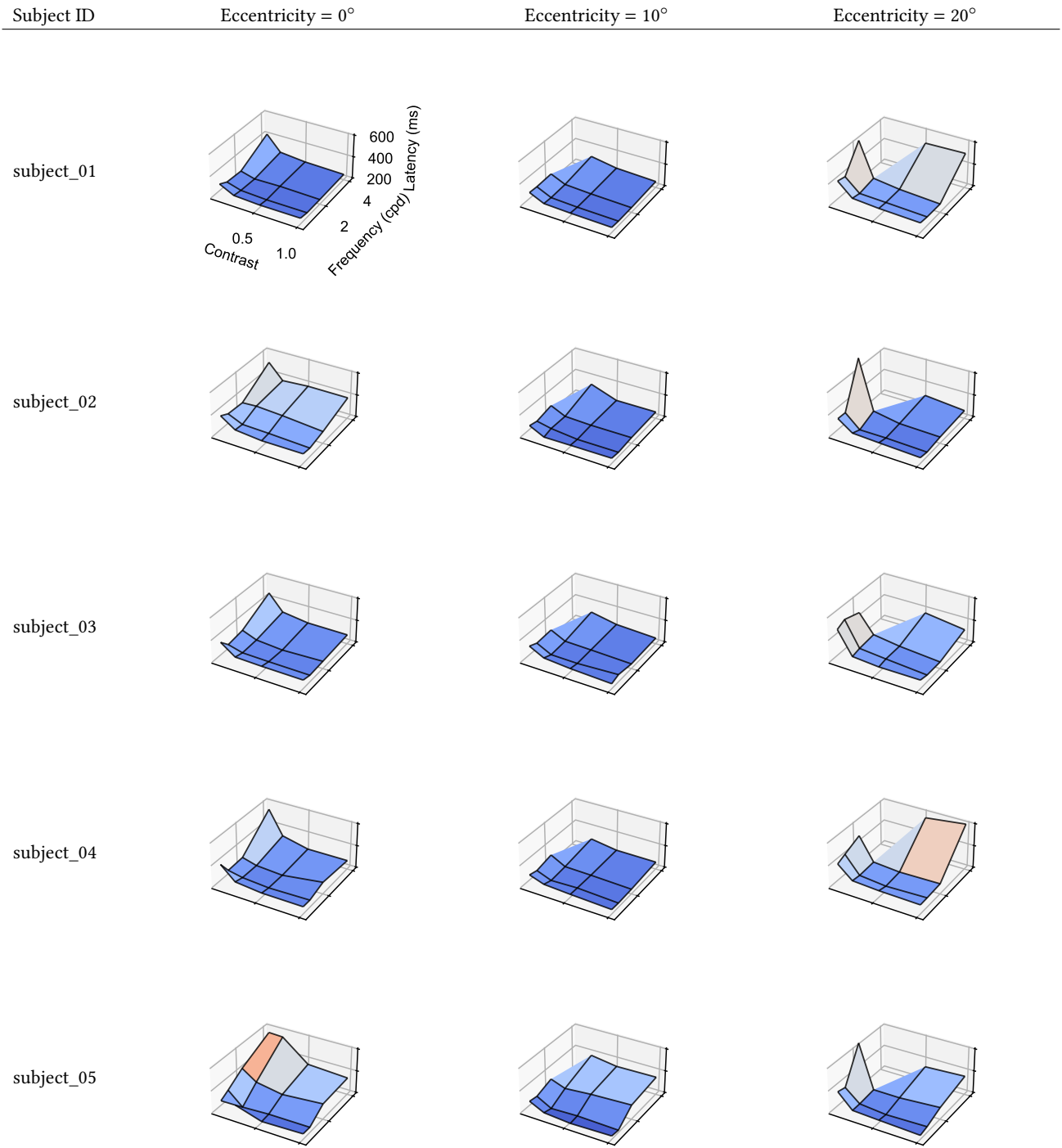

Fig. 10. Aggregated data of the pilot experiment. Each subject completed 50 repetitions for each of the 45 conditions across 10 blocks of the user study. Each vertex in these surfaces represent the mean saccade latency of 50 trials with the same condition for each subject.

## F PLOTS FOR ABLATION STUDY CONDITIONS (Section 6.1)

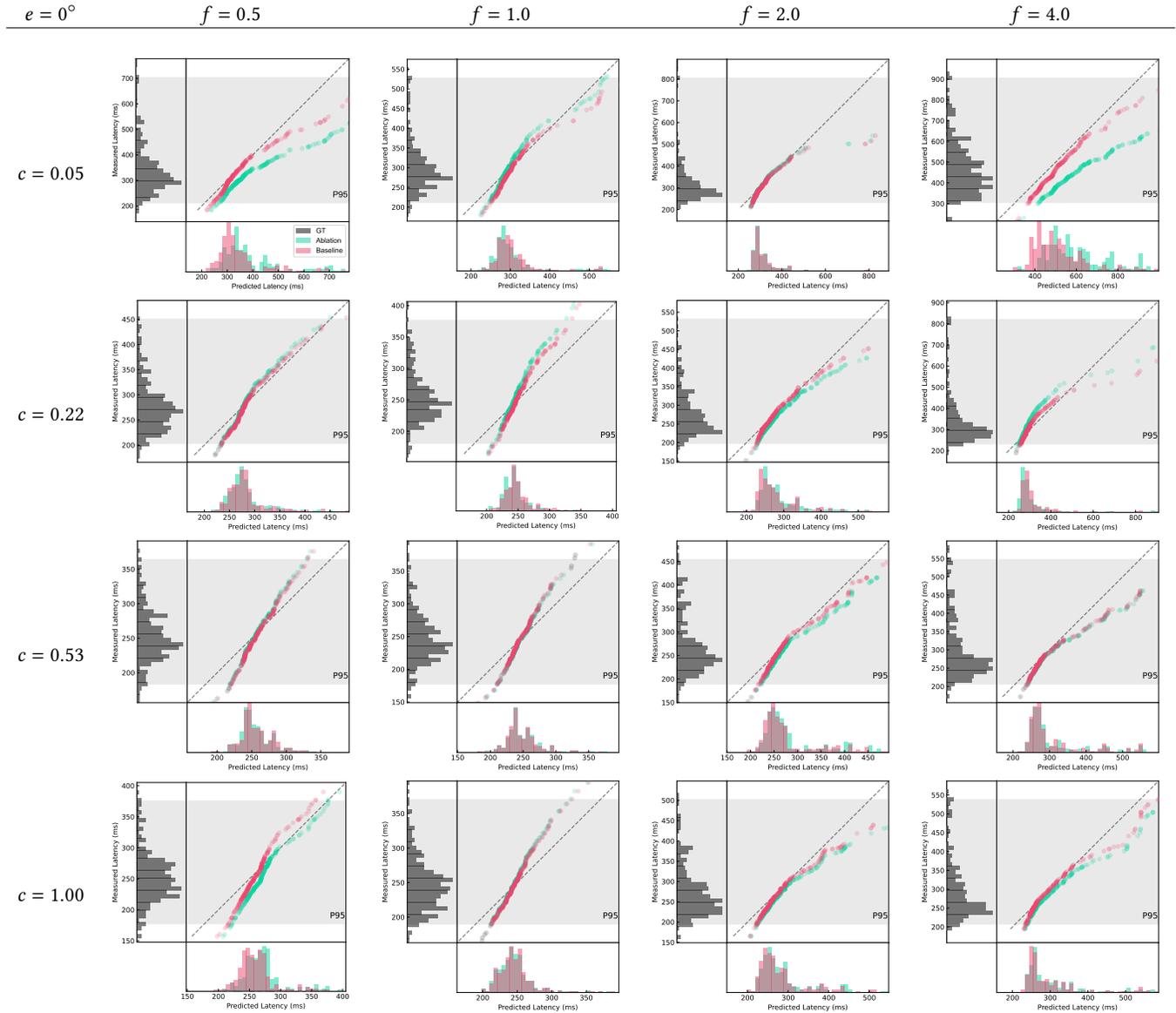

Fig. 11. Ablation study plots when any single condition is removed (as described in Section 6.1) from the training dataset where eccentricity $e = 0°$. By observing the corresponding model performance drop (i.e., stronger misalignment with the $y = x$ line), we visualize individual visual characteristic condition's contribution to the model. We observe that the distribution of latencies for some conditions cause a larger regression in the model's performance, such as the conditions ($c = 0.05, f = 0.5, e = 0°$) and ($c = 0.05, f = 4.0, e = 0°$). These regressions are caused by the fact that the model strongly relies on the data we collected for these specific conditions. Meanwhile, when conditions, such as ($c = 0.53, f = 1.0, e = 0°$), are removed for ablation the model is able to successfully interpolate their predictions, due to the abundance of neighbor conditions. To quantify the sizes of the regressions, we compute the MSE of ablated models against the ground truth data, and compare how much the error increased/decreased when compared to the full model. On average, the MSE of the ablated model regresses by as much as 50% when compared to the full model. However, the regression in performance is largely attributed by a few conditions which we mentioned above with the condition ($c = 0.05, f = 4.0, e = 0°$) exhibiting a 1100% increase in error. If we discount the extreme conditions, we observe that the median MSE regression is equal to 7%.

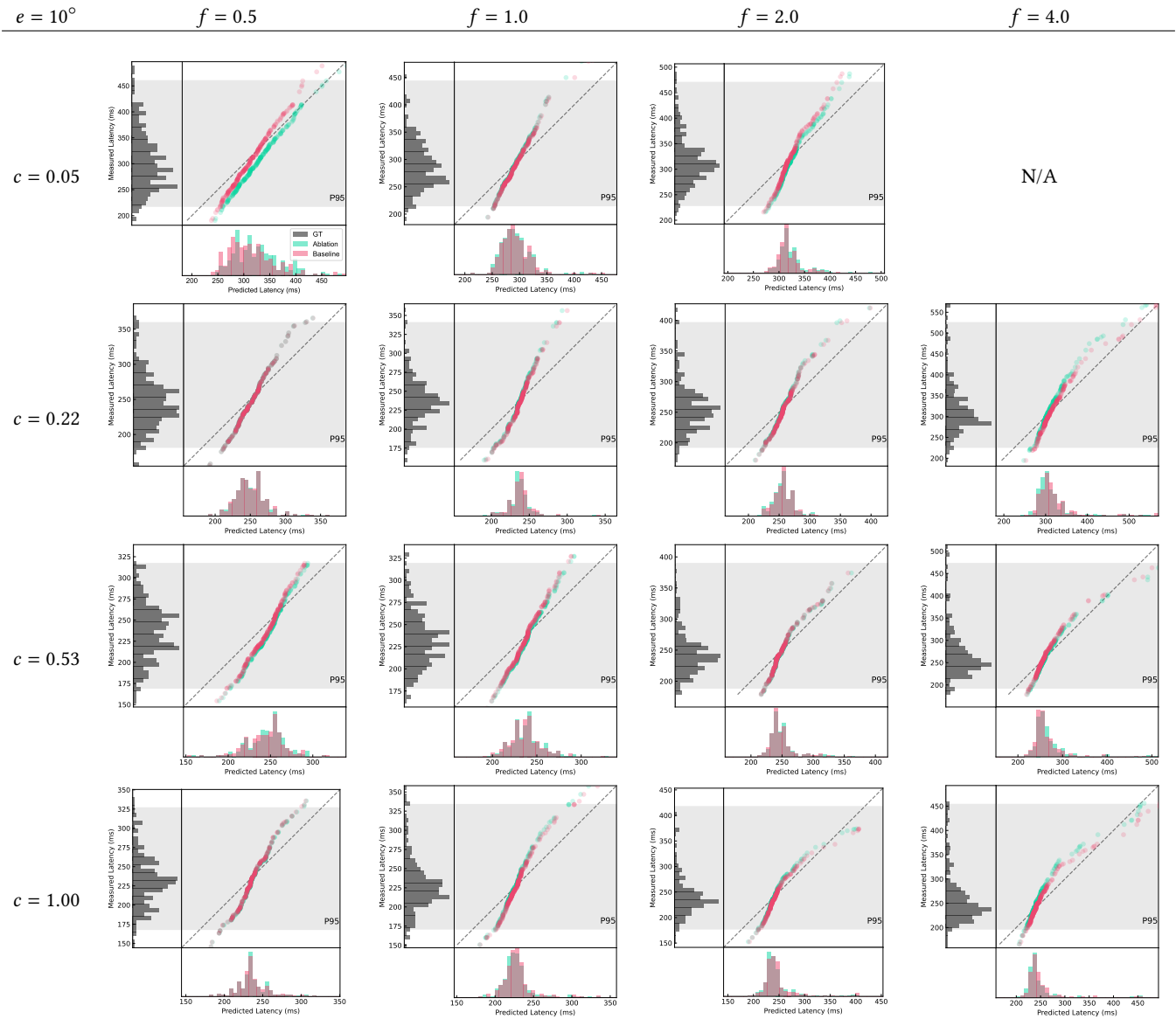

Fig. 12. Ablation study plots when any single condition is removed (as described in Section 6.1) from the training dataset where eccentricity $e = 10°$. See Figure 11, for further analysis.

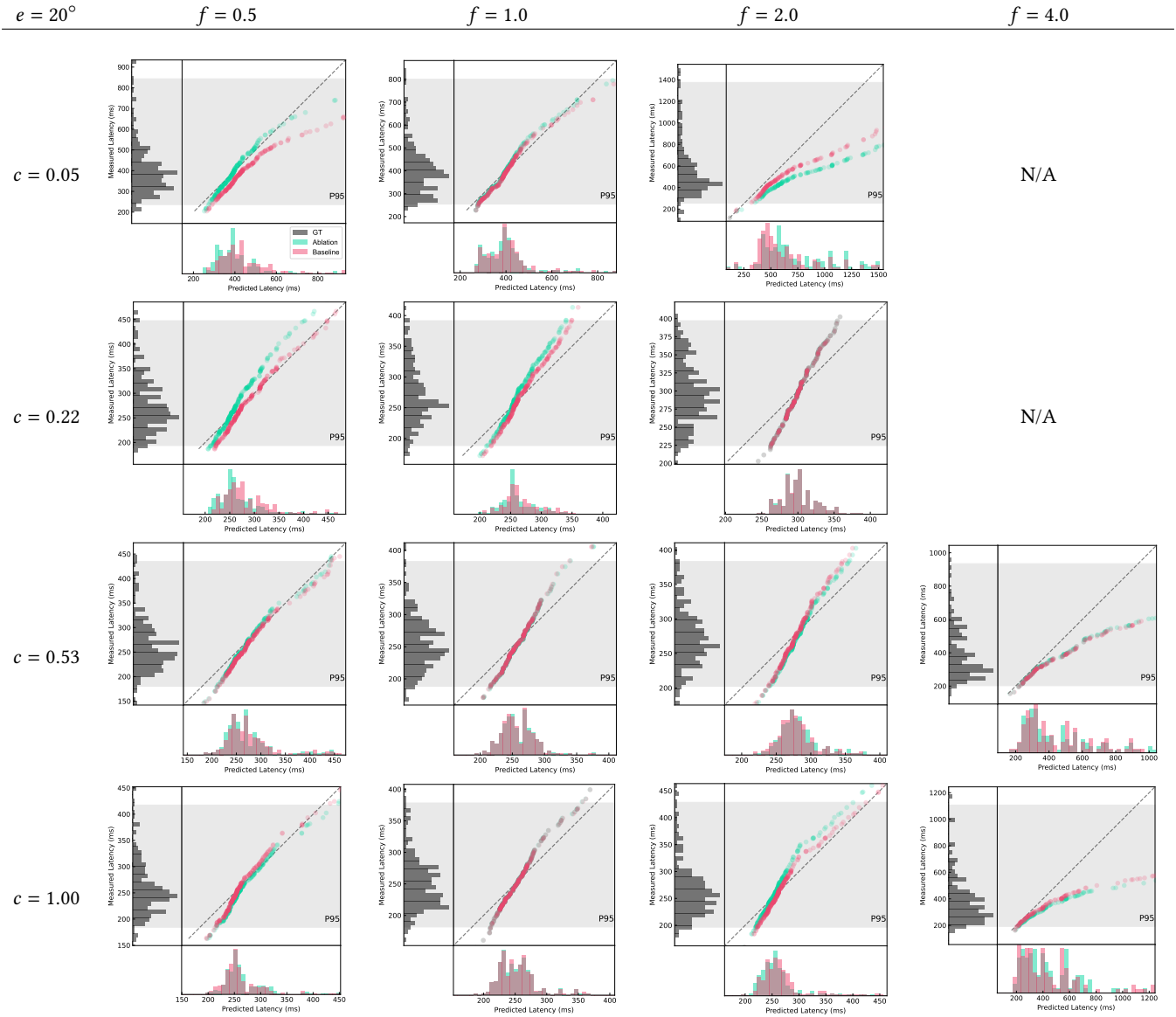

Fig. 13. Ablation study plots when any single condition is removed (as described in Section 6.1) from the training dataset where eccentricity $e = 20°$. See Figure 11, for further analysis.

# G    PLOTS FOR INDIVIDUAL PARTICIPANTS IN NATURAL TASKS (Section 6.2)

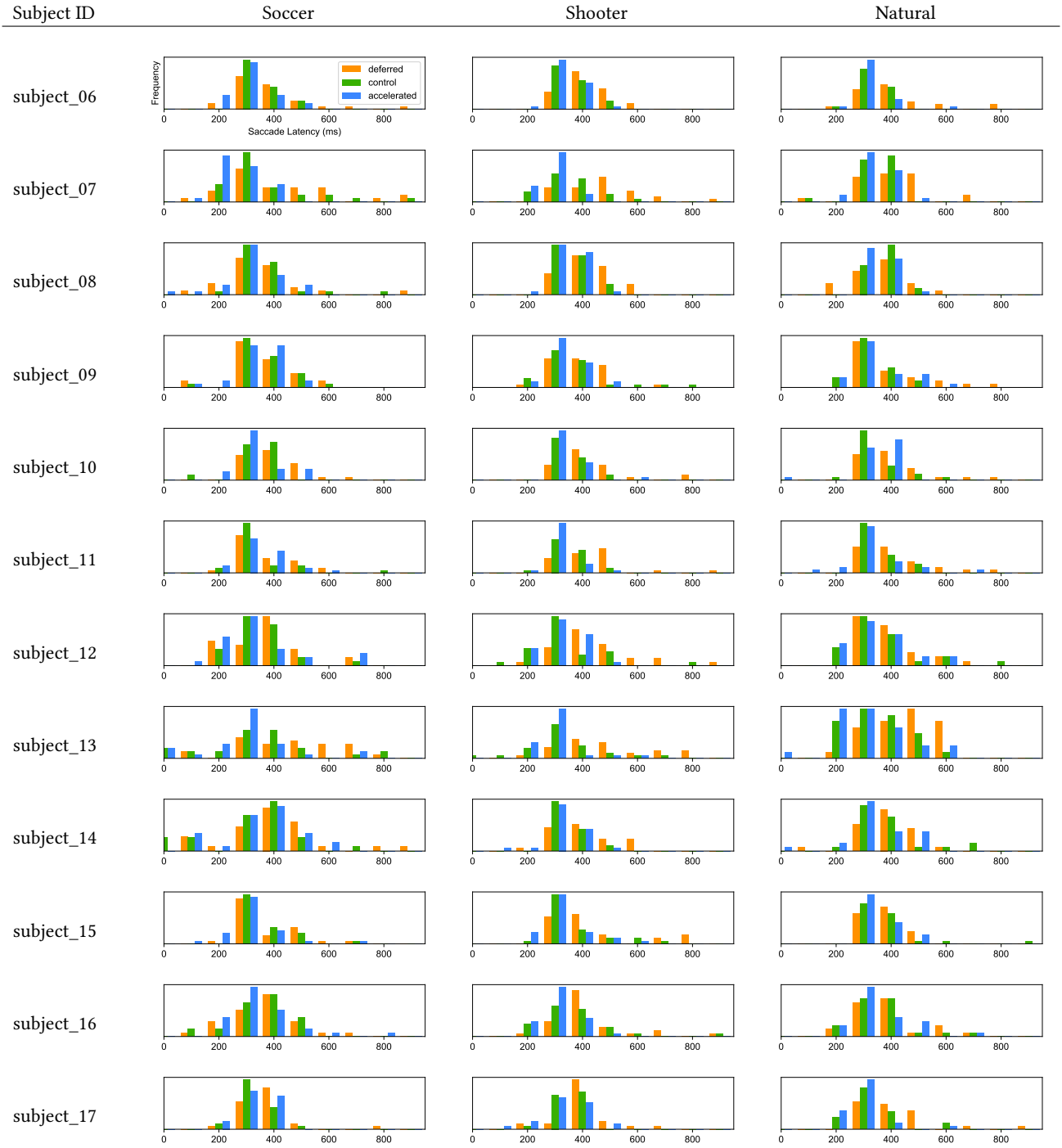

Fig. 14.  Saccade latency histograms for Figure 7. Each subject completed 51 trials for each condition, for each scene for a total of 459 trials. The latencies have been normalized to a common mean to enable quick comparisons between histograms.